\documentclass[prc,aps,twocolumn,showpacs]{revtex4}
\usepackage{graphicx}

\begin{document}

\title{Measurement of Inverse Pion Photoproduction \\
at Energies Spanning the $N(1440)$ Resonance}

\author{
A.~Shafi$^1$, S.~Prakhov$^2$, I.~I.~Strakovsky$^1$,
W.~J.~Briscoe$^1$\footnote[1]{Email: briscoe@gwu.edu},
B.~M.~K.~Nefkens$^2$, C.~E.~Allgower$^3$\footnote[2]
   {Present address: Midwest Proton Radiotherapy Institute,
   2425 N. Milo Sampson Lane, Bloomington, IN 47408, USA},
R.~A.~Arndt$^1$, V.~Bekrenev$^4$, C.~Bennhold$^1$, M.~Clajus$^2$,
J.~R.~Comfort$^5$, K.~Craig$^5$, D.~Grosnick$^6$,
D.~Isenhower$^7$, N.~Knecht$^8$, D.~D.~Koetke$^6$,
A.~Kulbardis$^4$, N.~Kozlenko$^4$, S.~Kruglov$^4$, G.~Lolos$^8$,
I.~Lopatin$^3$, D.~M.~Manley$^9$, R.~Manweiler$^6$,
A.~Maru\v{s}i\'{c}$^2$\footnote[3]{Present address:
   Collider-Accelerator Dept., Brookhaven National
   Laboratory, Upton, NY 11973.},
S.~McDonald$^2$\footnote[4]{Present address: TRIUMF,
   4004 Wesbrook Mall, Vancouver, B.C., Canada V6T
   2A3.},
J.~Olmsted$^9$\footnotemark[2], Z.~Papandreou$^8$,
D.~Peaslee$^{10}$, N.~Phaisangittisakul$^2$, J.~W.~Price$^2$,
A.~F.~Ramirez$^5$, M.~Sadler$^7$, H.~Spinka$^3$,
T.~D.~S.~Stanislaus$^6$, A.~Starostin$^{2,4}$,
H.~M.~Staudenmaier$^{11}$, I.~Supek$^{12}$,
W.~B.~Tippens$^2$\footnote[5]{Present address:
   Nuclear Physics Division, Department of Energy, 19901
   Germantown Road, Germantown, MD 20874-1290.},
and R.~L.~Workman$^1$ \\
\vspace*{0.1in}
(The Crystal Ball Collaboration)
\vspace*{0.1in}
}

%\address{
\affiliation{
$^1$The George Washington University, Washington, D.C. 20052-0001\\
$^2$University of California Los Angeles, Los Angeles, CA 90095-1547\\
$^3$Argonne National Laboratory, Argonne, IL 60439-4815\\
$^4$Petersburg Nuclear Physics Institute, Gatchina, Russia 188300\\
$^5$Arizona State University, Tempe, AZ 85287-1504\\
$^6$Valparaiso University, Valparaiso, IN 46383-6493\\
$^7$Abilene Christian University, Abilene, TX 79699-7963\\
$^8$University of Regina, Saskatchewan, Canada S4S OA2\\
$^9$Kent State University, Kent, OH 44242-0001\\
$^{10}$University of Maryland, College Park, Maryland 20742-4111\\
$^{11}$Universit\"at Karlsruhe, Karlsruhe, Germany 76128\\
$^{12}$Rudjer Boskovic Institute, Zagreb, Croatia 10002\\
}

\date{\today}

%%%%%%%%%%%%%%%%%%%%%%%%%%%%%%%%%%%%%%%%%%%%%%%%%%%%%%%%%%
%%%   Abstract
%%%%%%%%%%%%%%%%%%%%%%%%%%%%%%%%%%%%%%%%%%%%%%%%%%%%%%%%%%
\begin{abstract}
Differential cross sections for the process $\pi^-p\to\gamma n$
have been measured at Brookhaven National Laboratory's Alternating
Gradient Synchrotron with the Crystal Ball multiphoton
spectrometer. Measurements were made at 18 pion momenta from 238
to 748~MeV/$c$, corresponding to E$_{\gamma}$ for the inverse
reaction from 285 to 769~MeV.  The data have been used to evaluate
the $\gamma n$ multipoles in the vicinity of the $N(1440)$
resonance.  We compare our data and multipoles to previous
determinations. A new three-parameter SAID fit yields $36\pm7$
(GeV)$^{-1/2}\times 10^{-3}$ for the $A^n_{\frac{1}{2}}$ amplitude
of the $P_{11}$.
\end{abstract}


\maketitle

%\clearpage

%%%%%%%%%%%%%%%%%%%%%%%%%%%%%%%%%%%%%%%%%%%%%%%%%%%%%%%%%%
%%%   I. Introduction
%%%%%%%%%%%%%%%%%%%%%%%%%%%%%%%%%%%%%%%%%%%%%%%%%%%%%%%%%%
\section{Introduction}
\label{sec:intro}

The study of the light baryon resonances, in particular the
determination of their electromagnetic couplings, is undergoing a
resurgence driven by a stream of new, high-precision data
emanating from experimental programs at modern photon facilities.
The main properties of the $\Delta (1232)$ resonance, the mass,
pole value, mass splittings, width, and branching ratio for
different decay modes are now reasonably well known. In addition,
there are data on the deformation from a spherical shape, the
$E_2/M_1$ ratio, and the magnitude and sign of the magnetic dipole
moment of the $\Delta^{++}(1232)$. In comparison with this, the
properties of the lightest $N^{\ast}$ resonance, the $N(1440)$,
are much more uncertain. The Crystal Ball baryon-resonance program
at Brookhaven National Laboratory's (BNL) Alternating Gradient
Synchrotron (AGS) is providing much needed data on the $N(1440)$
resonance via new, precision measurements of $\gamma$, $\pi^0$,
and $2\pi^0$ production in $\pi^-p$ interactions. These data will
permit new, state-of-the-art coupled-channel analyses to be
performed. This article presents results of the $\pi^-p\to\gamma
n$ measurements.

The $N(1440)$, often called the Roper resonance, has the quantum
numbers of the nucleon ground state, $I, J^P = \frac{1}{2},
\frac{1}{2}^+$. The lightness of the mass is somewhat of a
surprise, as several models, including the relativized quark model
for baryons~\cite{cp92}, imply that the lowest mass states are the
$N(1520)\frac{3}{2}^-$ and $N(1535)\frac{1}{2}^-$. A better
understanding of the Roper resonance is particularly important as
its radiative decay width is larger than predicted. This has led
to a number of theoretical speculations concerning its underlying
structure.  The Roper resonance could be a radial excitation of
the nucleon~\cite{cp92} or a hybrid state consisting of three
quarks and a gluon~\cite{hybrid}. In an algebraic framework for
the description of baryons, Bijker, Iachello, and Leviatan studied
a collective string-like model to obtain masses and
electromagnetic couplings~\cite{bijk94}. Modern lattice-gauge
calculations with constrained curve-fitting techniques are also
now being used to study the Roper resonance~\cite{dong04}.
Recently, it was conjectured that the Roper resonance might be a
pentaquark state and a member of an antidecuplet~\cite{jaffe}. The
classification of the Roper resonance as an antidecuplet was
already proposed by Lovelace~\cite{lo65} back in 1965.
Donnachie~\cite{do67} suggested a simple test for this hypothesis.
By using \textit{U-spin} conservation one can easily show that the
radiative decay of the charged Roper resonance is not allowed:
$N(1440)\not\rightarrow\gamma p$; however, the radiative decay of
the neutral Roper resonance is allowed: $N(1440)\rightarrow\gamma
n$. The extracted $\gamma n$ and $\gamma p$ decay amplitudes are
not small (see Table~\ref{tbl1}), which does not support the
conjecture~\cite{jaffe} that the Roper resonance is a member of an
antidecuplet.

\begin{figure*}
\centering{
\includegraphics[height=0.45\textwidth, angle=90]{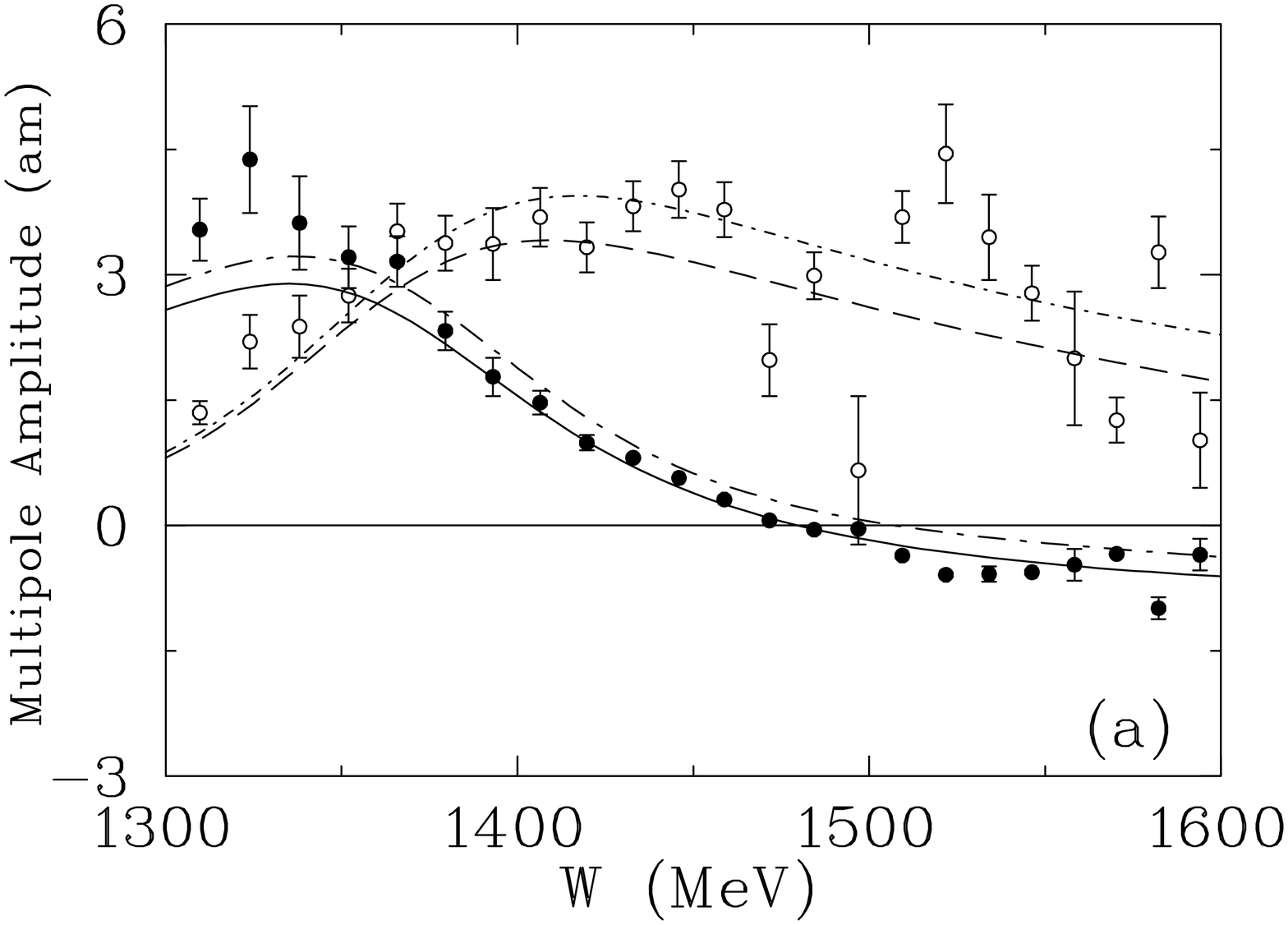}\hfill
\includegraphics[height=0.45\textwidth, angle=90]{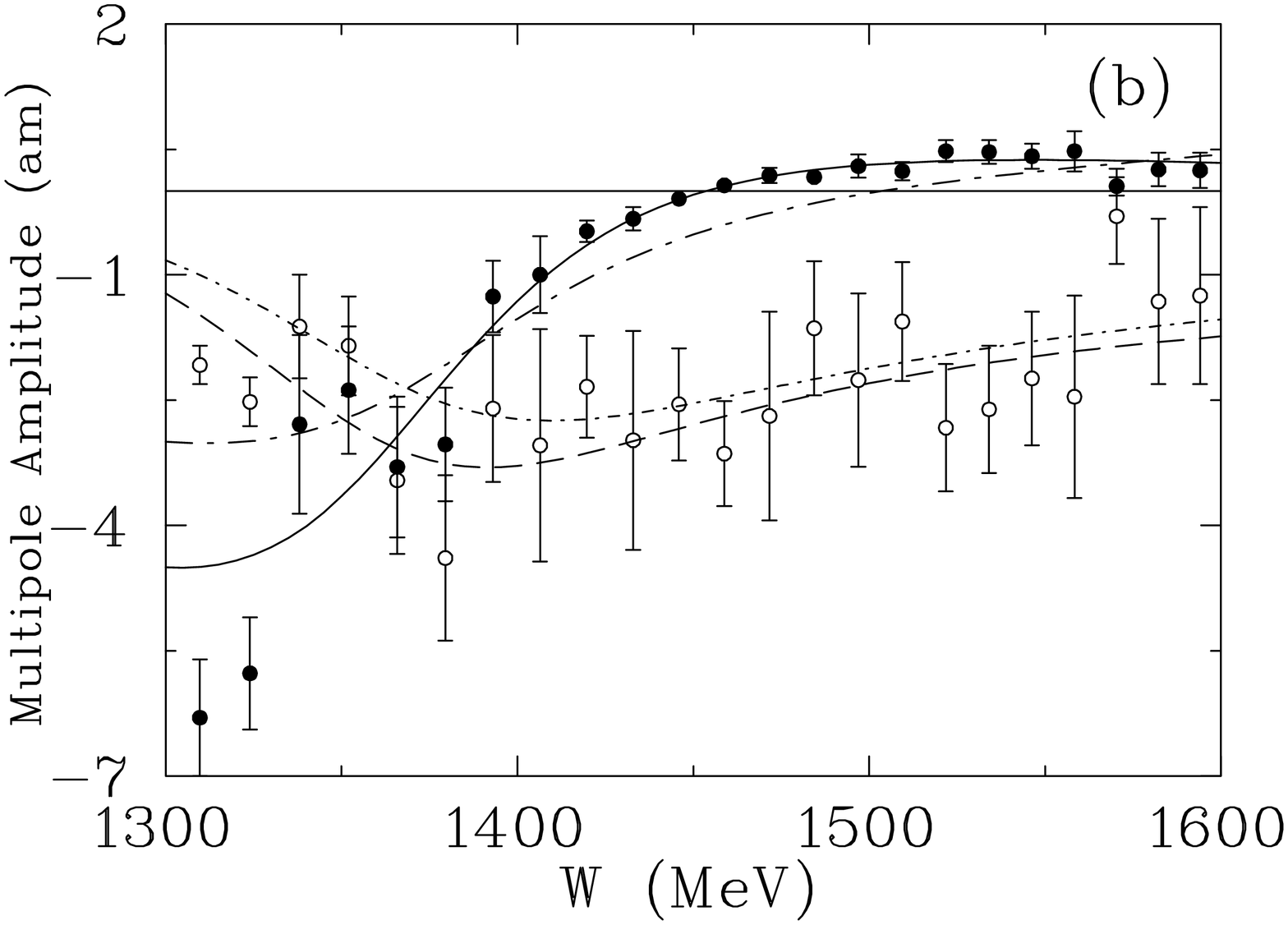}
}\caption{$M_{1-}^{1/2}$ multipoles in attometers
         (1~am $= 10^{-18}$m). Solid (dashed)
         curves give the real (imaginary) parts of
         amplitudes corresponding to the GW SM02
         solution~\protect\cite{GWpr}.  The real
         (imaginary) parts of GW single-energy
         solutions are plotted as filled (open)
         circles.  The MAID solution~\protect\cite{maid}
         is plotted with long dash-dotted (real part)
         and short dash-dotted (imaginary part) lines.
         Plotted are the multipole amplitudes (a)
         $\rm _pM_{1-}^{1/2}$ and (b) $\rm _nM_
         {1-}^{1/2}$.  The subscript p (n) denotes
         a proton (neutron) target.\label{fig:ampl}}
\end{figure*}

The radiative decay width of the charged Roper resonance is
readily extracted from $\pi^+$ and  $\pi^0$ photoproduction on a
proton. The radiative decay width of the neutral state may be
extracted from $\pi^-$ or $\pi^0$ photoproduction off a neutron,
which involves a bound neutron target (typically the deuteron) and
requires the use of a model-dependent nuclear correction. As a
result, our knowledge of neutral resonance decays is less precise
than of the charged ones. An example is given by the Roper
resonance photon-decay amplitudes listed in Table~\ref{tbl1}. The
PDG listings assign a 25\% uncertainty to the Roper resonance
$\gamma n$ amplitude, while a 6\% uncertainty is assigned to the
$\gamma p$ amplitude.  The associated photoproduction multipoles
are plotted in Fig.~\ref{fig:ampl}. Both the $\gamma p$ and
$\gamma n$ multipoles have sizable uncertainties at the energies
that correspond to the formation of the broad Roper resonance.

The existing photoproduction database contains a large set of
$\gamma n \rightarrow \pi^- p$ differential cross sections. Many
of these are old bremsstrahlung measurements with limited angular
coverage. In several cases, the systematic uncertainties have not
been quoted. An accurate treatment of final-state interaction
(FSI) effects for the pion photoproduction reactions on the
deuteron, $\gamma d\to\pi^- pp$ and $\gamma d\to\pi^0 np$, is
essential for the extraction of the spin-flip part of the
photoproduction amplitudes. In addition, the photon decay
amplitudes for the $N(1440)$ resonance, $A^p_{1/2}$ and
$A^n_{1/2}$, are similar in magnitude and opposite in sign,
suppressing the impulse-approximation contribution to the $\gamma
d \rightarrow \pi^-pp$ reaction.  As a result, diagrams involving
meson rescattering give a significant contribution to the full
amplitude.

%%%%%%%%%%%%%%%%%%%%%%%%%%%%%%%%%tbl.1
\begin{table}[h]
\caption{$N(1440)$ resonance couplings from a
         Breit-Wigner fit to the recent GW-SAID-2002
         single-energy solution [GW02]~\protect\cite{GWpr},
         the previous solution SM95 [VPI95]~\protect\cite{ar96},
         the analysis of Crawford and Morton
         [CM83]~\protect\cite{cm83},
         Crawford [CR01]~\protect\cite{cr01},
         Drechsel~\textit{et al.}
         [MAID98]~\protect\cite{maid98},
         [MAID03]~\protect\cite{maid},
         the coupled-channels fit of Niboh and Manley
         [KSU97]~\protect\cite{nib97}, GW-CC coupled channels
         calculation [BENN03]~\protect\cite{benn03}, the average
         of Feuster and Mosel [FM99]~\protect\cite{fm99},
         the average from the Particle Data Group
         [PDG02]~\protect\cite{PDG}, and quark model
         predictions by Capstick [CAP92]~\protect\cite{cp92}.
         Units are (GeV)$^{-1/2}\times 10^{-3}$. None of these
         include the results of this paper.} \label{tbl1}
\begin{tabular}{lccc}
%\tableline
\colrule
Resonance State & Reference   & $A^p_{1/2}$ & $A^n_{1/2}$\\
%\tableline
\colrule
$W_{Roper}\approx 1440$~MeV  & GW02         & $-$67$\pm$2 &  47$\pm$5  \\
$\Gamma_{\pi}/\Gamma\approx$ 0.65&VPI95& $-$63$\pm$5 &  45$\pm$15 \\
$\Gamma \approx 350$~MeV& CM83 & $-$69$\pm$18&  56$\pm$15 \\
                & CR01         & $-$88       &  $-$       \\
                & MAID98       & $-$71       &  60        \\
                & MAID03       & $-$77       &  52        \\
                & KSU97        & $-$81$\pm$6 &  65$\pm$12 \\
                & BENN03       & $-$81       &  59        \\
                & FM99         & $-$74       &  51        \\
                & PDG02        & $-$65$\pm$4 &  40$\pm$10 \\
                & CAP92        &     4       &  $-$6      \\
%\tableline
\colrule
\end{tabular}
\end{table}
%%%%%%%%%%%%%%%%%%%%%%%%%%%%%%%%%%%%%%%%
%%%%%%%%%%%%%%%%%%%%%%%%%%%%%%%%%%%%%%%%%
%%%%%%%%%%%%%%%%%%%%%%%%%%%%%%%%%

The radiative decay of the neutral Roper resonance can also be
obtained from the measurements of the inverse $\pi^-$
photoproduction reaction
\begin{eqnarray}
      \pi^-p\to\gamma n~,
\label{eqn:1}
\end{eqnarray}
which we will call REX for Radiative EXchange.  This process is
free from complications associated with the deuteron target.
However, the disadvantage of using this reaction is the high
background from the 5 to 500 times larger cross section for
\begin{eqnarray}
      \pi^-p\to\pi^0n\to\gamma\gamma n,
\label{eqn:2}
\end{eqnarray}
called CEX for Charge EXchange. The Crystal Ball (CB) multiphoton
spectrometer allows us to make a good measurement of the REX
reaction with the reliable subtraction of the CEX-reaction
background. These considerations motivated the measurement of
inverse pion photoproduction cross sections (E913)~\cite{Aziz} by
the CB Collaboration at the BNL-AGS.

An extensive set of measurements over the energy range associated
with the Roper resonance is essential to validate the existing
multipole analyses from which the radiative widths are extracted,
and to test the consistency of data that have been obtained with a
deuterium target. We report here on the determination of the
differential cross section for
\begin{eqnarray}
      \gamma n\to\pi^-p
\label{eqn:3}
\end{eqnarray}
from a measurement of the inverse reaction at 18 incident $\pi^-$
momenta from 238 to 748~MeV/$c$. This range corresponds to
$E_\gamma$ from 285 to 769~MeV for the inverse process, covering
the region most sensitive to the $N(1440)$ resonance; it
effectively doubles the database for the $\pi^-$ photoproduction
reaction. In Section~\ref{sec:distrib}, we discuss our analysis of
the differential cross sections.  The results of this experiment
are presented in Section~\ref{sec:dsg}.  We summarize our findings
in Section~\ref{sec:summ}.

%%%%%%%%%%%%%%%%%%%%%%%%%%%%%%%%%%%%%%%%%%%%%%%%%%%%%%%%%%
%%%   II. Experiment and Analysis
%%%%%%%%%%%%%%%%%%%%%%%%%%%%%%%%%%%%%%%%%%z%%%%%%%%%%%%%%%%
\section{Experimental set-up}
\label{sec:set-up}

Our measurements of $\pi^-p\to\gamma n$ were made at BNL with the
CB detector, which was installed in the AGS C6 beam line. The CB
consists of 672 NaI(Tl) crystals, each shaped like a truncated
triangular pyramid.  The crystals are optically isolated from
their neighbors and arranged in two hemispheres with an entrance
and exit tunnel for the beam and a spherical cavity in the center
for the target. The CB covers 93\% of $4\pi$ steradians.

The experiment was performed with a momentum-analyzed beam of
negative pions, incident on a 10-cm-long liquid hydrogen (LH$_2$)
target located in the center of the CB. The beam spread
$\sigma_p/p$ at the CB target was about 1\%. The uncertainty in
the mean momentum of the beam spectrum at the target center was
2--3~MeV/$c$.

The pulse height in each crystal was measured using a separate
ADC. For registering timing information, we used a TDC on every
minor triangle, which is a group of nine neighboring crystals. The
typical energy resolution for electromagnetic showers in the CB
was $\Delta E/E = 0.020/[E( \mathrm{GeV})]^{0.36}$. Showers were
measured with a resolution in $\theta$, the polar angle with
respect to the beam axis, of $\sigma _\theta = 2^\circ\textrm{--}
3^\circ$ for photon energies in the range 50--500~MeV, assuming
that the photons were produced in the center of the CB.  The
resolution in azimuthal angle $\phi$ is
$\sigma_\theta/\sin\theta$.

The CB event trigger required a beam trigger in coincidence with a
neutral event trigger, which included the requirement that the
total energy deposited in the CB crystals exceeded a certain
threshold. The beam trigger was a coincidence between three
scintillation counters located in the beam line upstream of the
CB.  The neutral event trigger required that the CB event trigger
signals were in anti-coincidence with signals from a barrel of
scintillation counters surrounding the target.

A more detailed description of the CB detector and the data
analyses can be found in Refs.~\cite{etalam,eta_slope,Aziz,2pi0n}.

%%%%%%%%%%%%%%%%%%%%%%%%%%%%%%%%%%%%%%%%%%%%%%%
\section{Data Analysis}
\label{sec:hand}

To select candidates for the $\pi^-p\to\gamma n$ reaction, we used
the neutral 1- and 2-cluster events, where we assumed that one of
the clusters was due to a photon electromagnetic shower in the CB.
A ``cluster'' is defined to be a group of neighboring crystals in
which energy is deposited from a single-photon electromagnetic
shower. The software threshold of the cluster energy was 14~MeV.
For a 1-cluster event, the missing particle was assumed to be the
neutron.  For 2-cluster events, one of the clusters was assumed to
come from a neutron interaction in the CB. The efficiency of the
CB for neutrons has been found in a separate test to vary from 0
to 30\% depending on the energy of the neutron~\cite{NIM01}. In
this experiment we used a lower cluster threshold (14~MeV) than in
the test run (20~MeV) which increased our maximum neutron
detection efficiency to 45\%.

Since the REX cross section is small, the handling of the
background is important. There are two kinds of background that
must be subtracted from the $\pi^-p\to\gamma n$ event sample.  The
principal background comes from the CEX reaction, when one of the
two photons from $\pi^0$ decay is not detected in the CB. Note
that the total cross section for the CEX reaction is about two
orders of magnitude larger than for the REX
reaction~\cite{GWpr,GWpiN3}. The effect of this background process
was estimated by determining the probability for Monte Carlo
simulated CEX events to be misidentified as $\pi^-p\to\gamma n$
candidates. The input needed for the simulation of this background
is the $\pi^-p\to\pi^0 n$ differential cross section that we have
measured at each beam momentum in the same experiment. The
fraction of events that are due to the CEX background depends
mainly on the ratio of the production rates for the two processes.
In the range of energies and angles reported in this article, this
fraction varies from 27\% at our lowest momentum, 238~MeV/$c$, to
59\% at our highest momentum, 748~MeV/$c$.

Other sources of background are due to processes that are not pion
interactions in the liquid hydrogen of the target. The main
contributions to this background are from beam pions that decayed
or scattered before reaching the target, or interacted in the
material surrounding the target. This background was investigated
in runs taken with an empty target. The fraction of events that
are due to the so-called ``empty-target" background is of the
order of a few percent, except at some extreme back angles.

All 1- and 2-cluster events were subjected to a kinematic fit to
test the hypothesis of process~(\ref{eqn:1}), while all 2- and
3-cluster events were tested for the hypothesis of
process~(\ref{eqn:2}). The kinematic fit has four main constraints
(4-C) based on energy and 3-momentum conservation. The hypothesis
for the CEX reaction has a fifth constraint that requires the
invariant mass of the two photons to be the known $\pi^0$-meson
mass. The measured parameters in the kinematic fit included five
for the beam particle (momentum, angles $\theta_x$ and $\theta_y$,
and position coordinates $x$ and $y$ in the target) and three for
each photon cluster (energy, angles $\theta$ and $\phi$).  When
the missing particle was the neutron, its energy and two angles
were free parameters in the fit. For the neutron detected in the
CB, the neutron angles were used as the measured parameters. In
the case of the CEX reaction, the $z$-coordinate of the primary
vertex was a free parameter in the kinematic fit. Since the
effective number of constraints is reduced by the number of free
parameters of the fit, for the CEX reaction we have a 1-C (3-C)
fit for 2-cluster (3-cluster) events. For the 1-cluster REX events
we could not use the $z$-coordinate of the vertex as a free
parameter in the fit because the effective number of constraints
would have been zero. To overcome this problem, the $z$-coordinate
was considered to be a "measured" parameter in the fit, with the
mean equal to the center of the target and the variance one third
of the target thickness. For the 1-cluster events, we had a 1-C
fit. The 2-cluster REX events have the neutron detected in the CB
and thus the $z$-coordinate can be a free parameter in a 2-C fit.

The confidence level (C.L.) of the kinematic fit was used to
select the REX candidates. The 1-cluster events that satisfied the
hypothesis above the 10\% C.L. (\textit{i.e.}, with a probability
greater than 10\%) were accepted as $\pi^-p\to\gamma n$
candidates.  The selection of 2-cluster REX events was performed
in two steps. In the first step, the neutron information was used
in the fit. This was necessary to suppress the large background
from the CEX reaction. Those events that satisfied the hypothesis
above the 1\% C.L. proceeded to the second step in which the
neutron information was omitted from the fit, and the event was
treated in the same way as the 1-cluster case. Since the kinematic
fit output is used for further analysis, this approach allows the
1- and 2-cluster events to have the same resolution for the photon
production angle. In further analysis, we considered only the sum
of 1- and 2-cluster events. This summation cancels problems
associated with the small difference between the real and
simulated events in the neutron response in the CB.

To select the CEX reaction events detected in the CB, we applied
just a 2\% C.L. criterion to hypothesis (\ref{eqn:2}) for 2- and
3-cluster events. Similar to the $\pi^-p\to\gamma n$ selection, we
added the 2- and 3-cluster events together. The only background
that had to be subtracted was the empty-target one.  The typical
fraction of events due to this background was about 5\%.

A Monte Carlo (MC) simulation of reaction~(\ref{eqn:1}) was
performed for each momentum according to the phase-space
distribution (\textit{i.e.}, with isotropic production angular
distribution). The CEX reaction was simulated twice, once
according to phase-space and once according to the shape of the
differential cross sections that were determined in this
experiment for each momentum.  The simulation was made for every
momentum by using the experimental beam-trigger events as input
for pion-beam distributions. The MC events were then propagated
through a full {\sc GEANT} (version 3.21)~\cite{GEANT} simulation
of the CB detector, folded with the CB resolutions and trigger
conditions, and analyzed in the same way as the experimental data.

The average detection efficiency for $\pi^-p\to\gamma n$ events
generated according to phase space varied between 57\% and 61\%
depending on the beam momentum and other experimental conditions.
The values are slightly less than the geometrical acceptance of
the CB for REX. The losses due to the exit hole in the CB are
aggravated by the forward boost of the final-state photon in the
laboratory system. Photon interactions in the beam pipe and in the
barrel of scintillation counters surrounding the target also
contribute to the loss of events. According to the simulation, the
average probability for a photon not to pass the neutral trigger
is about 6\%. Finally, some decrease in the acceptance occurs due
to the selection cuts used for the background suppression.

%%%%%%%%%%%%%%%%%%%%%%%%%%%%%%%%%%%%%%%%%%%
\section{The number of effective beam pions
and target protons}
\label{sec:b&t}

In addition to the determination of the number of initially
produced REX and CEX events, the cross-section calculation needs
the total number of beam pions, $N_{\pi^-}$, incident on the
target and the effective number of hydrogen atoms in the target.
The LH$_2$ target has a cylindrical shape along the beam direction
and has hemispherical endcaps. The maximum target thickness along
the beam axis is 10.57~cm. The hemisphere radius is 7.62~cm. The
effective hydrogen density for the LH$_2$ conditions calculated in
units of (mb$\cdot$cm)$^{-1}$ is
$\rho_{\mathrm{LH_2}}^{\mathrm{Eff}} = 4.248\times 10^{-5}$. The
effective number of hydrogen atoms is
$N_{\mathrm{LH_2}}^{\mathrm{Eff}} =
\rho_{\mathrm{LH_2}}^{\mathrm{Eff}}\times
l_{\mathrm{LH_2}}^{\mathrm{Eff}}$, where
$l_{\mathrm{LH_2}}^{\mathrm{Eff}}$ is the effective thickness of
the LH$_2$ target for the $\pi^-$ beam passing through the entire
target. This effective thickness was determined by a MC
simulation, where the real beam-trigger events were used for
calculating the average path length through the target.  Taking
into account the spatial distribution of the beam at different
momenta, the effective number of hydrogen atoms in the LH$_2$
target is $N_{\mathrm{LH_2}}^{\mathrm{Eff}} = (4.05\pm 0.08)\times
10^{-4}$~mb$^{-1}$.
%%%%%%%%%%%%%%%%%%%%%%%%%%%%%%%%%%%%%%%%%%%%%
\begin{figure}
\centering{
\includegraphics[height=0.45\textwidth, angle=90]{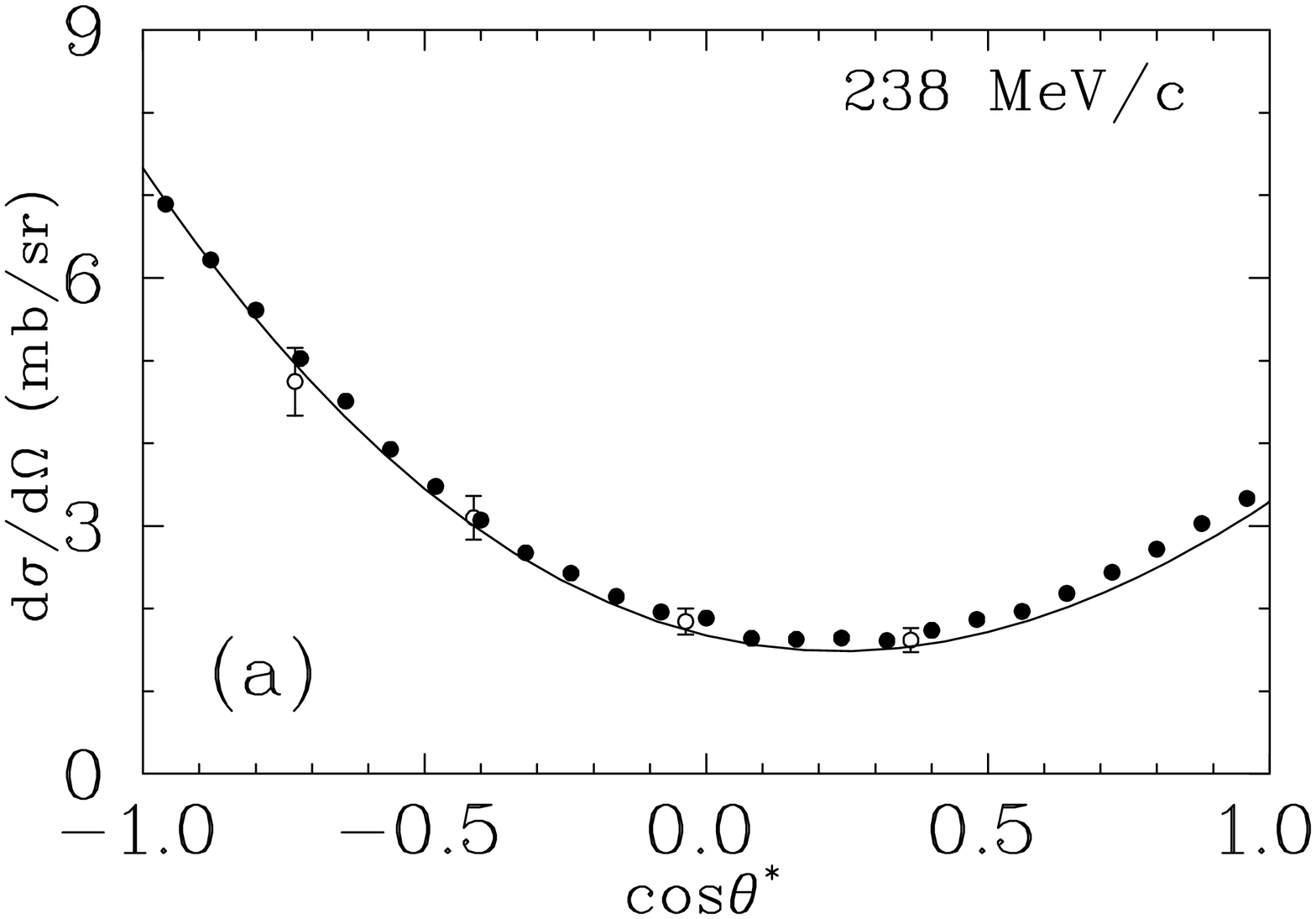}\vfill
\includegraphics[height=0.45\textwidth, angle=90]{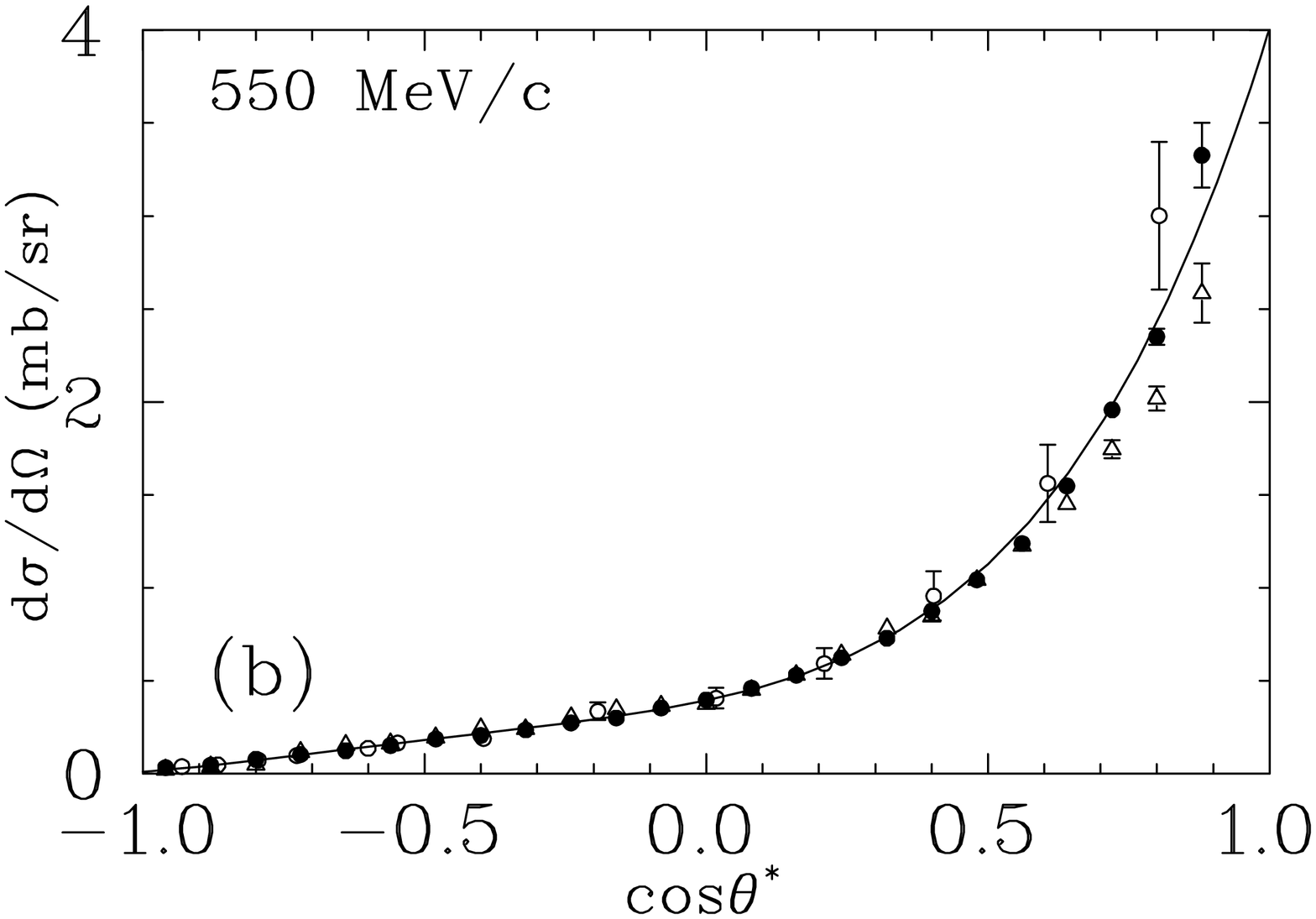}\vfill
\includegraphics[height=0.45\textwidth, angle=90]{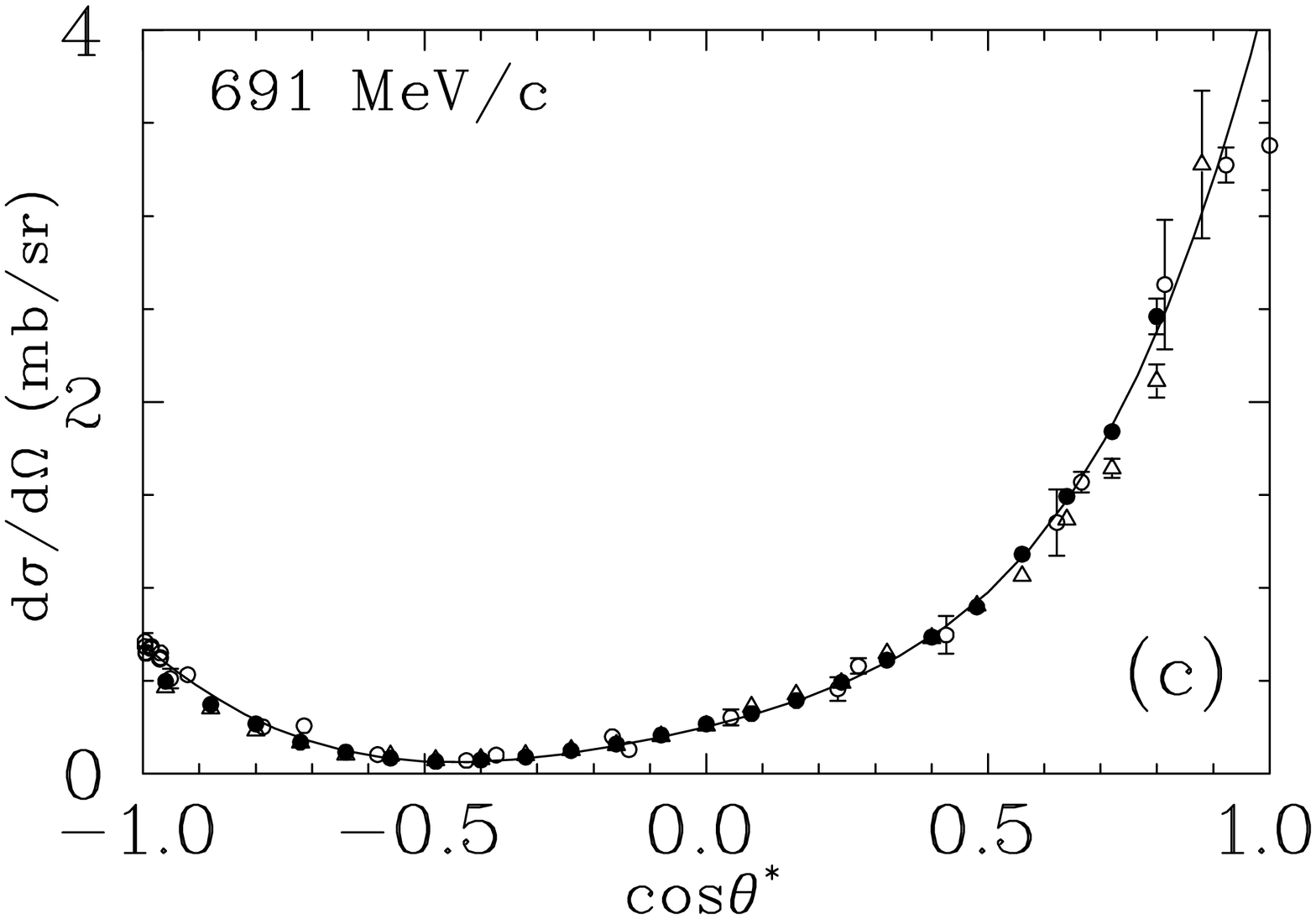}\vfill
}\caption{The differential cross section for
          $\pi^-p\to\pi^0n$ at an incident $\pi^-$
          momentum of (a) 238~MeV/$c$, (b) 550~MeV/$c$,
          and (c) 691~MeV/$c$. The LH$_2$ data (filled
          circles) at 550 and 691~MeV/$c$ are normalized
          to the central part of the CH$_2$ spectra (open
          triangles).  Solid lines show the GW SAID FA02
          predictions~\protect\cite{GWpiN3}.  Previous
          measurements~\protect\cite{CEXMES} are shown as open
          circles. \label{fig:norm_pi0n_pi656}}
\end{figure}
%%%%%%%%%%%%%%%%%%%%%%%%%%%%%%%%%%%%%%%%%%%%
%%%%%%%%%%%%%%%%%%%%%%%%%%%%%%%%%%%%%%%%%%%%
\begin{figure*}
\includegraphics[width=14.cm,height=14.cm,
bbllx=1.cm,bblly=1.cm,bburx=19.5cm,bbury=19.5cm]{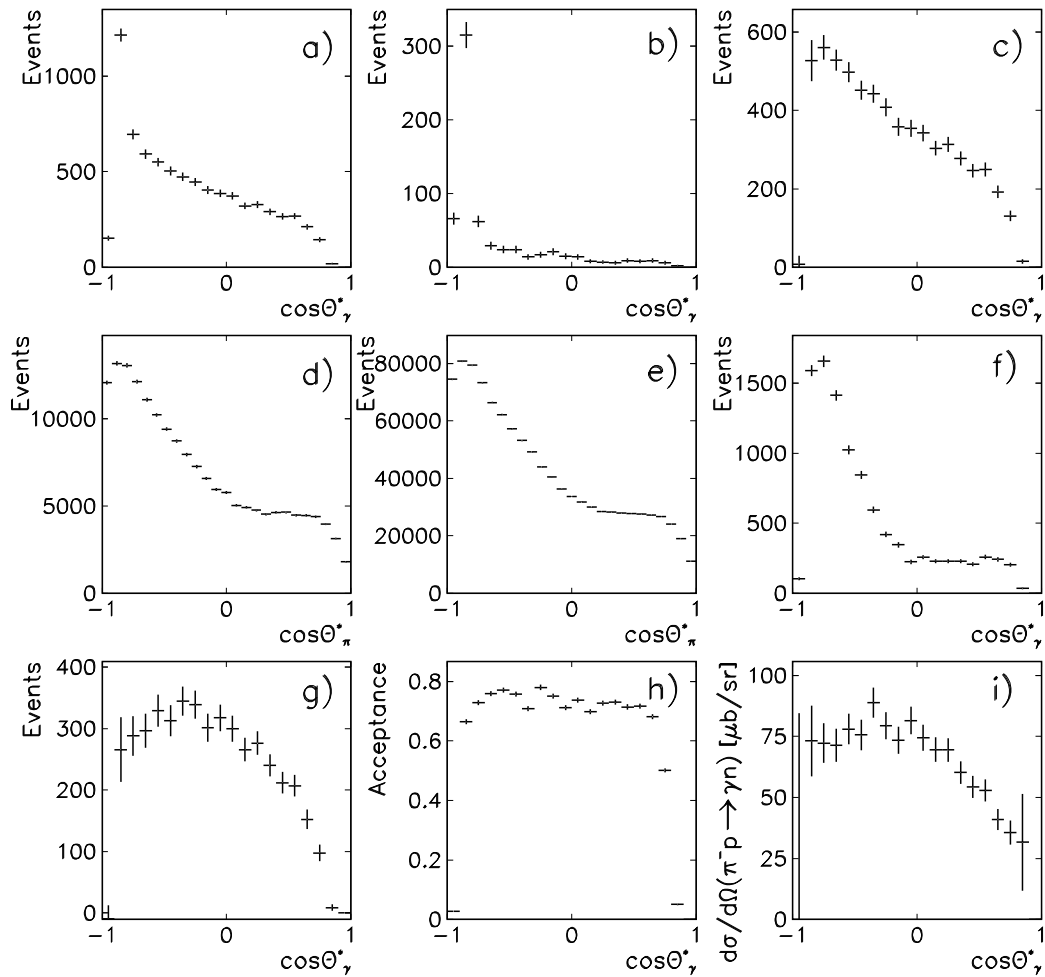}
\caption{The c.m. $\cos\theta^\ast$ distributions for the
         production angle of the photon from $\pi^-p\to\gamma n$
         and the $\pi^0$ from $\pi^-p\to\pi^0n$ at $p_{\pi^-}
         = 238$~MeV/$c$: ~(a) the experimental candidates for
         $\pi^-p\to\gamma n$; ~(b) the empty-target events
         selected as $\pi^-p\to\gamma n$; ~(c) the REX
         candidates after the empty-target background
         subtraction; ~(d) the experimental CEX events; ~(e)
         the MC simulation for CEX events; ~(f) the CEX
         MC-simulation events misidentified as REX; ~(g) the
         REX events after both the empty-target and the
         CEX-background subtractions; ~(h) acceptance for
         $\pi^-p\to\gamma n$; and ~(i) the differential cross
         section for $\pi^-p\to\gamma n$.
} \label{fig:gn_pi238}
\end{figure*}
%%%%%%%%%%%%%%%%%%%%%%%%%%%%%%%%%%%%%%%%%
%%%%%%%%%%%%%%%%%%%%%%%%%%%%%%%%%%%%%%%%%
\begin{figure*}
\includegraphics[width=14.cm,height=14.cm,bbllx=1.cm,bblly=1.cm,
bburx=19.5cm,bbury=19.5cm]{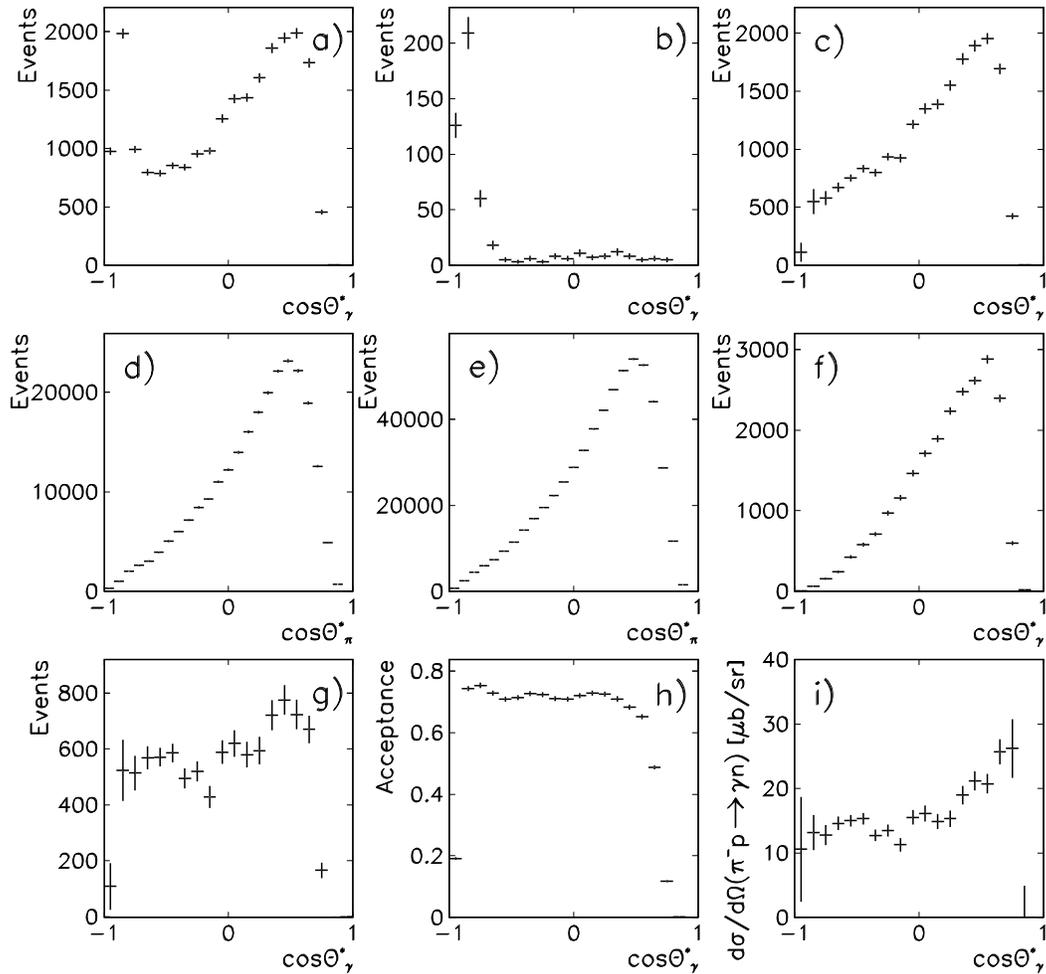}
\caption{Same as Fig.~\protect{\ref{fig:gn_pi238}} but for
         $p_{\pi^-} = 550$~MeV/$c$.
\label{fig:gn_pi550}}
\end{figure*}
%%%%%%%%%%%%%%%%%%%%%%%%%%%%%%%%%%%%%%%%%%
%%%%%%%%%%%%%%%%%%%%%%%%%%%%%%%%%%%%%%%%%
\begin{figure*}
\includegraphics[width=14.cm,height=14.cm,bbllx=1.cm,bblly=1.cm,
bburx=19.5cm,bbury=19.5cm]{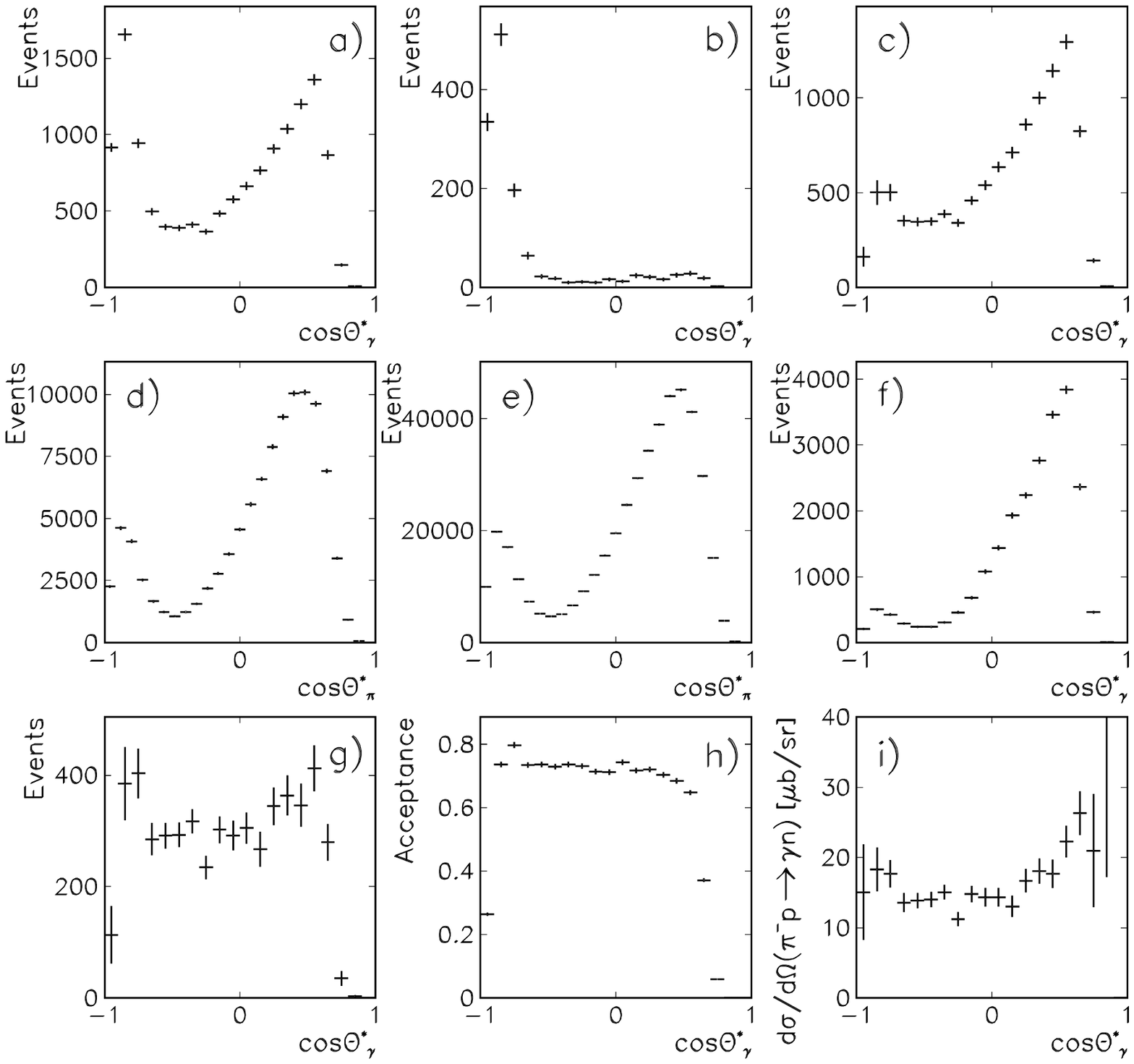}
\caption{Same as Fig.~\protect{\ref{fig:gn_pi238}} but for
         $p_{\pi^-} = 691$~MeV/$c$.
\label{fig:gn_pi691}}
\end{figure*}
%%%%%%%%%%%%%%%%%%%%%%%%%%%%%%%%%%%%%%%%%%

The calculation of $N_{\pi^-}$ involves several corrections that
take into account the scattering and the decay of pions, and also
the contamination of the pion beams by muons and
electrons~\cite{E890}. The decay and scattering of the beam pions
were taken into account by simulation. The real beam-trigger
events were used as input for this simulation. The trajectory
information for the beam particles was measured by the drift
chambers located in the beam line.  For beam momenta below
350~MeV/$c$, the beam contamination by muons and electrons was
measured by time-of-flight (TOF). There are several measurements
of the CEX reaction in this energy range; a comparison of our CEX
results with the existing data showed good agreement~\cite{Mike}.
At all momenta, the electron contamination was also investigated
by using a \v{C}erenkov counter located in the beam downstream of
the CB. However, our CEX results based on the \v{C}erenkov counter
information fell below the existing data. Since we relied entirely
on the \v{C}erenkov counter above 350~MeV/$c$, another
normalization method was used in our analysis.

The normalization of all LH$_2$ data sets with beam momenta above
350~MeV/$c$ was made by remeasuring the CEX reaction with solid
CH$_2$ targets. For the CH$_2$ measurements, the beam control was
optimized to diminish uncertainties in the number of pions
incident on the target. In Fig.~\ref{fig:norm_pi0n_pi656}(a), we
show our LH$_2$ results for the CEX reaction at $p_{\pi^-} =
238$~MeV/$c$. These results are obtained by using the beam
information taken with the LH$_2$ target. In
Fig.~\ref{fig:norm_pi0n_pi656}(b,c), we illustrate the
normalization of the LH$_2$ data at $p_{\pi^-}= 550$ and
691~MeV/$c$ to the corresponding CH$_2$ spectra. To exclude the
effect of low acceptance in the forward angles, we used only the
angular region $\cos\theta^{\ast}_{\pi^0} < 0.6$ for the
normalization.  More details about the beam normalization of the
LH$_2$ data can be found in Ref.~\cite{2pi0n}.  In
Fig.~\ref{fig:norm_pi0n_pi656} we also show the corresponding
SAID~\cite{GWpiN3} FA02 partial-wave analysis (PWA) of the
existing data.

%%%%%%%%%%%%%%%%%%%%%%%%%%%%%%%%%%%%%%%%%%%%%%%%%%%%
\section{Determination of the $\pi^-p\to\gamma n$
differential cross sections}
\label{sec:distrib}

Figure~\ref{fig:gn_pi238} illustrates our procedure for
determining the differential cross section of
reaction~(\ref{eqn:1}) for our lowest beam momentum, 238~MeV/$c$.
Each distribution in the figure is shown as a function of
$\cos\theta^\ast$, where $\theta^{\ast}$ is the angle between the
photon (or $\pi^0$) direction and the beam direction in the $\pi^-
p$ center-of-mass (c.m.) system.  In Fig.~\ref{fig:gn_pi238}(a),
one can see the experimental distribution for all events selected
as $\pi^-p\to\gamma n$ candidates.  The empty-target background
distribution is shown in Fig.~\ref{fig:gn_pi238}(b).  It can be
seen that the largest contamination occurs at backward angles due
to the decay of pions in the beam. The distribution remaining
after the empty-target background subtraction is shown in
Fig.~\ref{fig:gn_pi238}(c). The subtraction of this background was
made with a weight equal to the ratio of the number of incident
pions for the full and the empty targets, respectively. This ratio
varies from 2 to 5 depending on the relative beam on target of the
full- and empty-target runs at each momentum. In
Fig.~\ref{fig:gn_pi238}(d), we show the experimental distribution
of the events selected as the CEX reaction after the empty-target
background subtraction. In Fig.~\ref{fig:gn_pi238}(e), one can see
the CEX distribution reconstructed for $2\times 10^6$
$\pi^-p\to\pi^0n$ events simulated according to our differential
cross section obtained for this reaction at the given momentum.
The CEX background in the $\pi^-p\to\gamma n$ events is shown in
Fig.~\ref{fig:gn_pi238}(f). It was obtained from the simulated CEX
events that survived the selection criteria for
reaction~(\ref{eqn:1}). This background looks somewhat similar to
the CEX distribution itself; however, the average probability for
the CEX events to be misidentified as $\pi^-p\to\gamma n$ is about
0.8\%. The $\cos\theta^\ast_\gamma$ distribution remaining after
both the empty-target and the CEX background subtractions is shown
Fig.~\ref{fig:gn_pi238}(g). The CEX background subtraction was
made with a weight equal to the ratio of the CEX events
reconstructed from the data and from the MC simulation. The
acceptance for the $\pi^-p\to\gamma n$ reaction at $p_{\pi^-} =
238$~MeV/$c$ is shown in Fig.~\ref{fig:gn_pi238}(h).  This
acceptance is about 75\% for the central angles and drops in the
forward and backward direction. Finally, the resulting
differential cross section of reaction~(\ref{eqn:1}) at $p_{\pi^-}
= 238$~MeV/$c$ is shown in Fig.~\ref{fig:gn_pi238}(i) in units of
$\mu$b/sr. To calculate the acceptance-corrected
$\cos\theta^\ast_\gamma$ spectrum in these units, the number of
events in a particular bin of the spectrum was multiplied by the
factor $1000/(2\pi\times N_{\pi^-}\times N_{\mathrm{LH_2}}
^{\mathrm{Eff}}\times\Delta\cos\theta^\ast_\gamma)$, where
$\Delta\cos\theta^\ast_\gamma$ is the bin width. The uncertainties
in all distributions shown in Fig.~\ref{fig:gn_pi238} are
statistical only.

The same procedure was carried out at each of our 18 beam momenta.
To illustrate the determination of differential cross sections at
higher momenta, we show in Figs.~\ref{fig:gn_pi550} and
\ref{fig:gn_pi691} similar distributions for $p_{\pi^-} = 550$ and
691~MeV/$c$.  Note that increasing the beam momentum results in an
increased probability for the CEX events to be misidentified as
$\pi^-p\to\gamma n$; at $p_{\pi^-} = 550$ and 691~MeV/$c$ it is
about 5\%.

The main sources of experimental uncertainty are: (i) the
background subtraction, (ii) the acceptance correction, and (iii)
the normalization procedure.

The uncertainty in the background subtraction has two components:
the subtraction of CEX events that pass the REX event selection
and the beam-related background. The uncertainty in the CEX cross
section was estimated to be about 5\% which includes the addition
in quadrature of the uncertainty of $\sim4\%$ based on the
comparison of the SAID PWA results for the CEX reaction and our
measured differential cross section, an uncertainty of $\sim3\%$
for central values of $\cos\theta^{\ast}$ of the CEX angular
distribution, and an uncertainty of $\sim2\%$ in the normalization
of the LH$_2$ data relative to the CH$_2$ data. The final
uncertainty due to this factor varies with the magnitude of the
point by point subtraction of CEX events that are misidentified as
REX events; it is largest at backward and/or forward angles
(depending on beam energy) where the subtraction is large
(yielding an uncertainty of a few percent) and is smallest at
central angles where the subtraction is small (yielding and
uncertainty of a fraction of a percent). Similarly, the
uncertainty in the relative normalization of the the beam-related
empty-target background is $\sim3\%$ and has a larger effect at
backward angles (several percent) than at central to forward
angles (about a percent). The uncertainty in the REX differential
cross section due to these factors is determined bin by bin by
obtaining cross sections in the standard manner and then with each
background increased by its uncertainty.

The acceptance is flat over most of the angular range with an
uncertainty at the $\sim1\%$ level at all but the most forward
angles, where the value for the acceptance drops off rapidly and
the relative uncertainty approaches a few percent. The
uncertainties due to the background subtractions and acceptance
corrections are angle dependent and are included in the
uncertainties reported for the points listed in Tables II - IV.

The overall normalization uncertainty in our REX results of
$\sim5\%$ is mainly due to the estimated uncertainties described
above for the measured CEX cross section at the central values for
its angular distribution, the estimated uncertainty in the SAID
CEX cross section, and uncertainty in the normalization of the
LH$_2$ data relative to the CH$_2$ data. This total systematic
uncertainty is not included in the figures and tables.

%%%%%%%%%%%%%%%%%%%%%%%%%%%%%%%%%tbl.2
\begin{table*}
\caption{Differential cross section for $\gamma n\to\pi^-p$ (in
$\mu$b/sr) as a function of center-of-mass scattering angle and
pion laboratory momentum (top row of header), and photon energy
(bottom row of header).  The quoted uncertainties are statistical
and include the angle-dependent uncertainties due to the
subtraction process and acceptance corrections. The total overall
systematic uncertainty is about 5\%. These are described in the
text.}
\begin{turnpage}
\begin{ruledtabular}
\label{tbl2}
\begin{tabular}{ccccccc}
%\tableline
\colrule $\cos\theta^{\ast}_\pi$
     &238$\pm$3~MeV/$c$&271$\pm$3~MeV/$c$&298$\pm$3~MeV/$c$&322$\pm$3~MeV/$c$&355$\pm$4~MeV/$c$&373$\pm$4~MeV/$c$\\
     &285$\pm$3~MeV  &313$\pm$3~MeV  &338$\pm$3~MeV  &359$\pm$3~MeV  &390$\pm$4~MeV  &407$\pm$4~MeV\\
%\tableline
\colrule
$-$0.85&25.5$\pm$5.6&29.9$\pm$5.2&15.9$\pm$3.9&17.5$\pm$3.8& 8.5$\pm$2.8&13.0$\pm$1.9\\
$-$0.75&25.2$\pm$3.1&29.2$\pm$3.1&18.3$\pm$2.4&14.9$\pm$2.3&10.1$\pm$1.8& 8.9$\pm$1.0\\
$-$0.65&24.9$\pm$2.6&28.8$\pm$2.8&23.4$\pm$2.2&19.5$\pm$2.0&11.4$\pm$1.5&10.3$\pm$0.8\\
$-$0.55&27.2$\pm$2.3&28.3$\pm$2.6&23.2$\pm$2.0&15.8$\pm$1.8&12.0$\pm$1.4&10.2$\pm$0.7\\
$-$0.45&26.4$\pm$2.2&24.4$\pm$2.4&23.4$\pm$1.9&17.7$\pm$1.7&12.9$\pm$1.4& 9.6$\pm$0.7\\
$-$0.35&31.0$\pm$2.2&30.5$\pm$2.3&19.7$\pm$1.8&17.9$\pm$1.8&11.1$\pm$1.3& 9.9$\pm$0.6\\
$-$0.25&27.7$\pm$1.9&27.4$\pm$2.1&21.1$\pm$1.7&17.8$\pm$1.7&10.4$\pm$1.3& 9.3$\pm$0.6\\
$-$0.15&25.6$\pm$2.0&23.8$\pm$2.0&21.0$\pm$1.6&17.0$\pm$1.5&12.6$\pm$1.4&11.1$\pm$0.6\\
$-$0.05&28.4$\pm$2.0&25.9$\pm$2.0&24.4$\pm$1.7&14.4$\pm$1.5&12.3$\pm$1.3&10.4$\pm$0.7\\
   0.05&25.9$\pm$1.9&26.5$\pm$1.9&24.7$\pm$1.7&17.4$\pm$1.6&12.1$\pm$1.4&11.7$\pm$0.7\\
   0.15&24.3$\pm$1.8&28.1$\pm$2.1&23.4$\pm$1.7&19.9$\pm$1.6&13.5$\pm$1.5&11.7$\pm$0.8\\
   0.25&24.2$\pm$1.7&23.1$\pm$2.0&23.2$\pm$1.7&17.1$\pm$1.7&11.8$\pm$1.5&11.6$\pm$0.9\\
   0.35&21.0$\pm$1.6&22.5$\pm$1.9&19.6$\pm$1.6&20.9$\pm$1.8&15.4$\pm$1.7&12.0$\pm$0.9\\
   0.45&18.9$\pm$1.6&19.5$\pm$1.8&20.6$\pm$1.8&17.1$\pm$1.8&14.6$\pm$1.7&13.7$\pm$1.1\\
   0.55&18.4$\pm$1.6&17.6$\pm$1.9&19.4$\pm$1.7&16.8$\pm$1.8&15.1$\pm$1.9&13.4$\pm$1.2\\
   0.65&14.3$\pm$1.5&15.0$\pm$1.8&15.6$\pm$1.9&16.3$\pm$2.0&14.1$\pm$2.1&14.2$\pm$1.4\\
   0.75&12.4$\pm$1.7&15.1$\pm$2.3&18.3$\pm$2.6&21.4$\pm$2.8&15.7$\pm$2.8&11.9$\pm$1.7\\
%\tableline
\colrule
\end{tabular}
\end{ruledtabular}
\end{turnpage}
\end{table*}
%%%%%%%%%%%%%%%%%%%%%%%%%%%%%%%%%tbl.3
\begin{table*}
\caption{Continuation of Table~\protect\ref{tbl2}.}
\begin{ruledtabular}
\label{tbl3}
\begin{tabular}{ccccccc}
%\tableline
\colrule $\cos\theta^{\ast}_\pi$
     &404$\pm$4~MeV/$c$&472$\pm$5~MeV/$c$&550$\pm$5~MeV/$c$&656$\pm$6~MeV/$c$&668$\pm$6~MeV/$c$&678$\pm$6~MeV/$c$\\
     &436$\pm$4~MeV  &501$\pm$5~MeV  &576$\pm$5~MeV  &679$\pm$6~MeV  &691$\pm$6~MeV  &700$\pm$6~MeV\\
%\tableline
\colrule
$-$0.85& 7.3$\pm$1.2&10.6$\pm$1.4& 6.0$\pm$1.3& 6.2$\pm$0.9& 8.2$\pm$1.1& 4.9$\pm$1.4\\
$-$0.75& 7.1$\pm$0.6& 5.7$\pm$0.9& 5.8$\pm$0.7& 6.4$\pm$0.6& 7.2$\pm$0.7& 7.2$\pm$0.8\\
$-$0.65& 7.2$\pm$0.6& 6.0$\pm$0.7& 6.6$\pm$0.5& 6.2$\pm$0.4& 6.4$\pm$0.5& 6.4$\pm$0.6\\
$-$0.55& 7.8$\pm$0.6& 6.4$\pm$0.6& 6.8$\pm$0.4& 6.5$\pm$0.3& 5.8$\pm$0.5& 6.4$\pm$0.5\\
$-$0.45& 8.1$\pm$0.5& 5.2$\pm$0.6& 7.0$\pm$0.4& 6.8$\pm$0.3& 6.4$\pm$0.4& 6.2$\pm$0.5\\
$-$0.35& 7.5$\pm$0.5& 6.4$\pm$0.6& 5.8$\pm$0.5& 5.6$\pm$0.4& 6.9$\pm$0.5& 6.4$\pm$0.5\\
$-$0.25& 8.1$\pm$0.6& 6.7$\pm$0.6& 6.1$\pm$0.5& 6.0$\pm$0.4& 6.3$\pm$0.5& 5.9$\pm$0.5\\
$-$0.15& 7.8$\pm$0.6& 5.3$\pm$0.7& 5.1$\pm$0.6& 6.1$\pm$0.6& 6.4$\pm$0.6& 6.5$\pm$0.6\\
$-$0.05& 8.9$\pm$0.6& 7.5$\pm$0.8& 7.1$\pm$0.6& 7.2$\pm$0.5& 7.3$\pm$0.7& 7.8$\pm$0.6\\
   0.05& 9.9$\pm$0.7& 8.4$\pm$0.9& 7.3$\pm$0.7& 7.0$\pm$0.6& 7.4$\pm$0.7& 7.6$\pm$0.7\\
   0.15&10.3$\pm$0.8& 8.4$\pm$0.9& 6.8$\pm$0.7& 6.7$\pm$0.7& 7.1$\pm$0.8& 7.3$\pm$0.8\\
   0.25& 9.9$\pm$0.8& 8.2$\pm$1.0& 7.0$\pm$0.8& 8.0$\pm$0.7& 8.8$\pm$0.9& 8.8$\pm$0.9\\
   0.35&11.2$\pm$0.9& 8.6$\pm$1.1& 8.6$\pm$0.9& 8.9$\pm$0.9& 8.4$\pm$1.0&10.2$\pm$1.1\\
   0.45&11.2$\pm$1.1&10.3$\pm$1.2& 9.6$\pm$1.0& 9.2$\pm$1.0&10.4$\pm$1.2& 9.4$\pm$1.2\\
   0.55&12.2$\pm$1.2& 9.4$\pm$1.4& 9.4$\pm$1.1&10.2$\pm$1.2&10.1$\pm$1.3&10.5$\pm$1.4\\
   0.65&12.2$\pm$1.3&11.4$\pm$1.5&11.7$\pm$1.3&12.6$\pm$1.4& 9.2$\pm$1.7&10.1$\pm$1.7\\
   0.75&11.9$\pm$1.8&15.2$\pm$2.2&11.9$\pm$2.3&11.5$\pm$2.5&13.5$\pm$3.5&12.3$\pm$3.2\\
%\tableline
\colrule
\end{tabular}
\end{ruledtabular}
\end{table*}
%%%%%%%%%%%%%%%%%%%%%%%%%%%%%%%%%tbl.4
\begin{table*}
\caption{Continuation of Table~\protect\ref{tbl2}.}
\begin{ruledtabular}
\label{tbl4}
\begin{tabular}{ccccccc}
%\tableline
\colrule $\cos\theta^{\ast}_\pi$
     &691$\pm$6~MeV/$c$&704$\pm$7~MeV/$c$&719$\pm$7~MeV/$c$&727$\pm$7~MeV/$c$&733$\pm$6~MeV/$c$&748$\pm$7~MeV/$c$\\
     &713$\pm$6~MeV  &726$\pm$7~MeV  &740$\pm$7~MeV  &748$\pm$7~MeV  &754$\pm$6~MeV  &769$\pm$7~MeV\\
%\tableline
\colrule
$-$0.85& 8.6$\pm$1.7& 7.7$\pm$0.9& 9.4$\pm$1.4& 6.5$\pm$1.2& 7.7$\pm$1.9& 4.5$\pm$1.4\\
$-$0.75& 8.3$\pm$1.0& 8.7$\pm$0.6& 6.9$\pm$0.9& 6.5$\pm$0.8& 3.9$\pm$1.3& 5.6$\pm$0.9\\
$-$0.65& 6.4$\pm$0.7& 6.9$\pm$0.5& 6.3$\pm$0.7& 6.6$\pm$0.6& 5.0$\pm$0.9& 6.1$\pm$0.7\\
$-$0.55& 6.5$\pm$0.5& 6.2$\pm$0.4& 6.3$\pm$0.6& 5.8$\pm$0.5& 5.6$\pm$0.6& 4.8$\pm$0.6\\
$-$0.45& 6.6$\pm$0.5& 6.1$\pm$0.4& 6.2$\pm$0.5& 5.3$\pm$0.5& 4.9$\pm$0.5& 4.7$\pm$0.6\\
$-$0.35& 7.0$\pm$0.5& 6.0$\pm$0.4& 6.1$\pm$0.5& 4.6$\pm$0.5& 5.2$\pm$0.5& 4.4$\pm$0.5\\
$-$0.25& 5.2$\pm$0.5& 5.9$\pm$0.4& 5.8$\pm$0.5& 5.8$\pm$0.5& 5.7$\pm$0.5& 3.9$\pm$0.5\\
$-$0.15& 6.9$\pm$0.6& 5.5$\pm$0.4& 6.5$\pm$0.6& 6.0$\pm$0.5& 4.9$\pm$0.6& 4.2$\pm$0.6\\
$-$0.05& 6.7$\pm$0.7& 7.4$\pm$0.6& 5.0$\pm$0.7& 6.6$\pm$0.6& 5.7$\pm$0.7& 5.3$\pm$0.7\\
   0.05& 6.7$\pm$0.7& 6.9$\pm$0.6& 6.8$\pm$0.8& 6.1$\pm$0.7& 5.6$\pm$0.8& 5.0$\pm$0.8\\
   0.15& 6.1$\pm$0.9& 7.0$\pm$0.7& 6.0$\pm$0.9& 7.2$\pm$0.8& 5.9$\pm$0.9& 6.3$\pm$0.9\\
   0.25& 7.8$\pm$1.0& 8.0$\pm$0.9& 6.5$\pm$1.0& 7.1$\pm$1.0& 6.5$\pm$1.1& 6.4$\pm$1.0\\
   0.35& 8.5$\pm$1.1& 9.5$\pm$1.1&10.7$\pm$1.3& 8.1$\pm$1.2& 7.3$\pm$1.3& 7.3$\pm$1.2\\
   0.45& 8.3$\pm$1.3& 8.8$\pm$1.2& 9.3$\pm$1.4& 7.9$\pm$1.4& 7.3$\pm$1.5& 6.5$\pm$1.4\\
   0.55&10.4$\pm$1.5& 8.9$\pm$1.5& 5.9$\pm$1.7& 7.3$\pm$1.6& 7.5$\pm$1.6& 7.0$\pm$1.6\\
   0.65&12.3$\pm$1.9&10.2$\pm$1.8& 8.6$\pm$2.2& 7.6$\pm$2.0& 7.6$\pm$2.2& 7.4$\pm$2.1\\
%\tableline
\colrule
\end{tabular}
\end{ruledtabular}
\end{table*}
\printtables

%%%%%%%%%%%%%%%%%%%%%%%%%%%%%%%%%%%%%%%%%%%%%%%%%%%%%
%%%  III. Results for gamma n -> pi- p
%%%%%%%%%%%%%%%%%%%%%%%%%%%%%%%%%%%%%%%%%%%%%%%%%%%%%
\section{Results for $\gamma n\to\pi^-p$}
\label{sec:dsg}

The complete collection of the results of our experiment on
$d\sigma / d\Omega (\pi^-p\rightarrow\gamma n)$ are given in the
thesis by Shafi~\cite{Aziz}. Here, we present our data converted
to the inverse process.  This facilitates the comparison with the
numerous data sets that exist for $\pi^+$ and $\pi^0$
photoproduction on a hydrogen target. Assuming time-reversal
invariance, the radiative $\pi^-p$ capture is related to $\pi^-$
photoproduction on the neutron via the detailed balance relation:
\begin{eqnarray}
  d\sigma(\gamma n\rightarrow\pi^-p) = DB
  ~d\sigma(\pi^-p\rightarrow\gamma n) ,
\label{eqn:4}
\end{eqnarray}
where $DB = \frac{1}{2}(\frac{p^{\ast}_\pi} {p^{\ast}_\gamma})^2$
is the detailed balance factor, the number $\frac{1}{2}$ is the
spin-factor weight for the process, and $p^{\ast}$ is the momentum
in the center of mass.

\begin{figure*}
\centering{
\includegraphics[height=0.45\textwidth, angle=90]{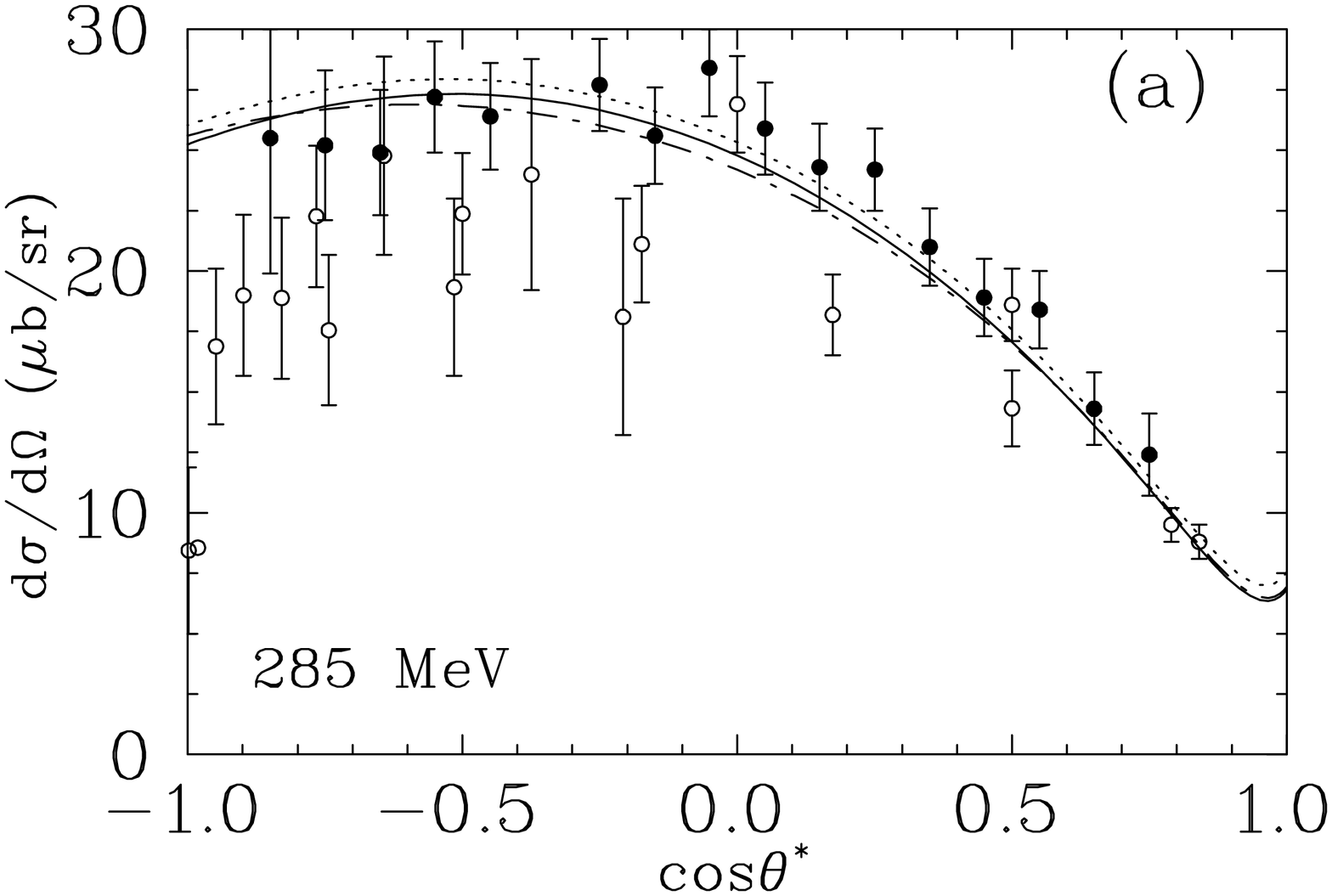}\hfill
\includegraphics[height=0.45\textwidth, angle=90]{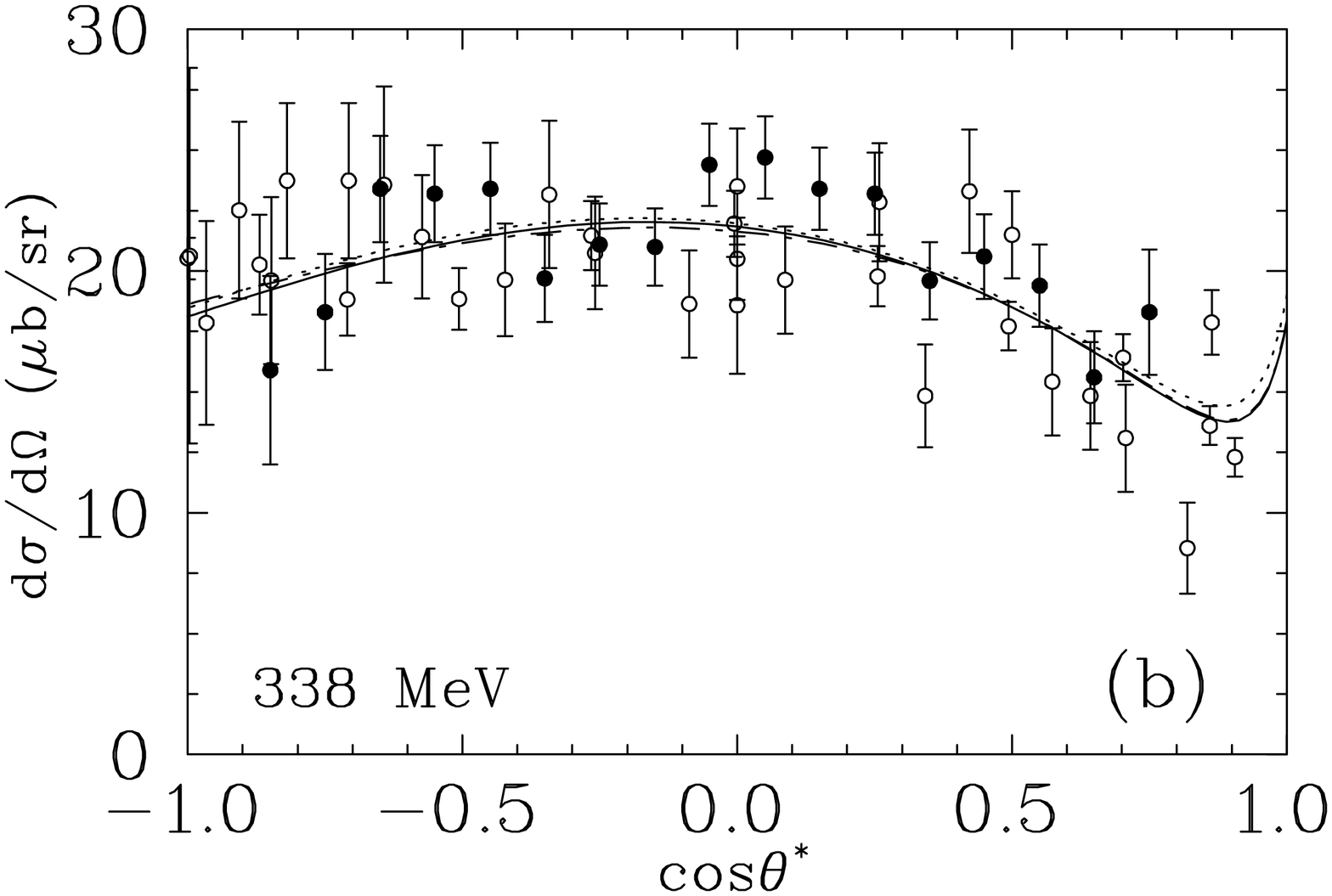}
\includegraphics[height=0.45\textwidth, angle=90]{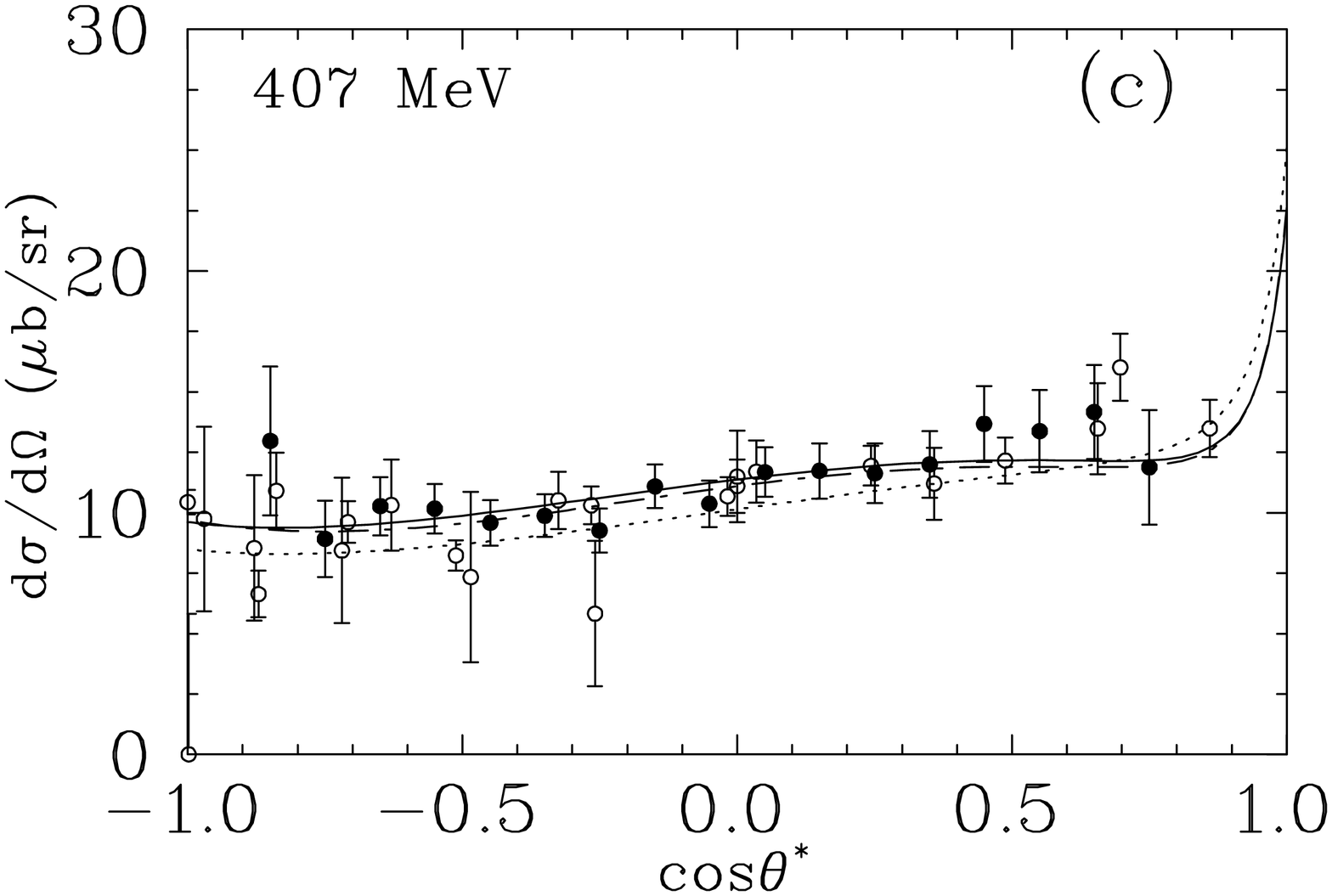}\hfill
\includegraphics[height=0.45\textwidth, angle=90]{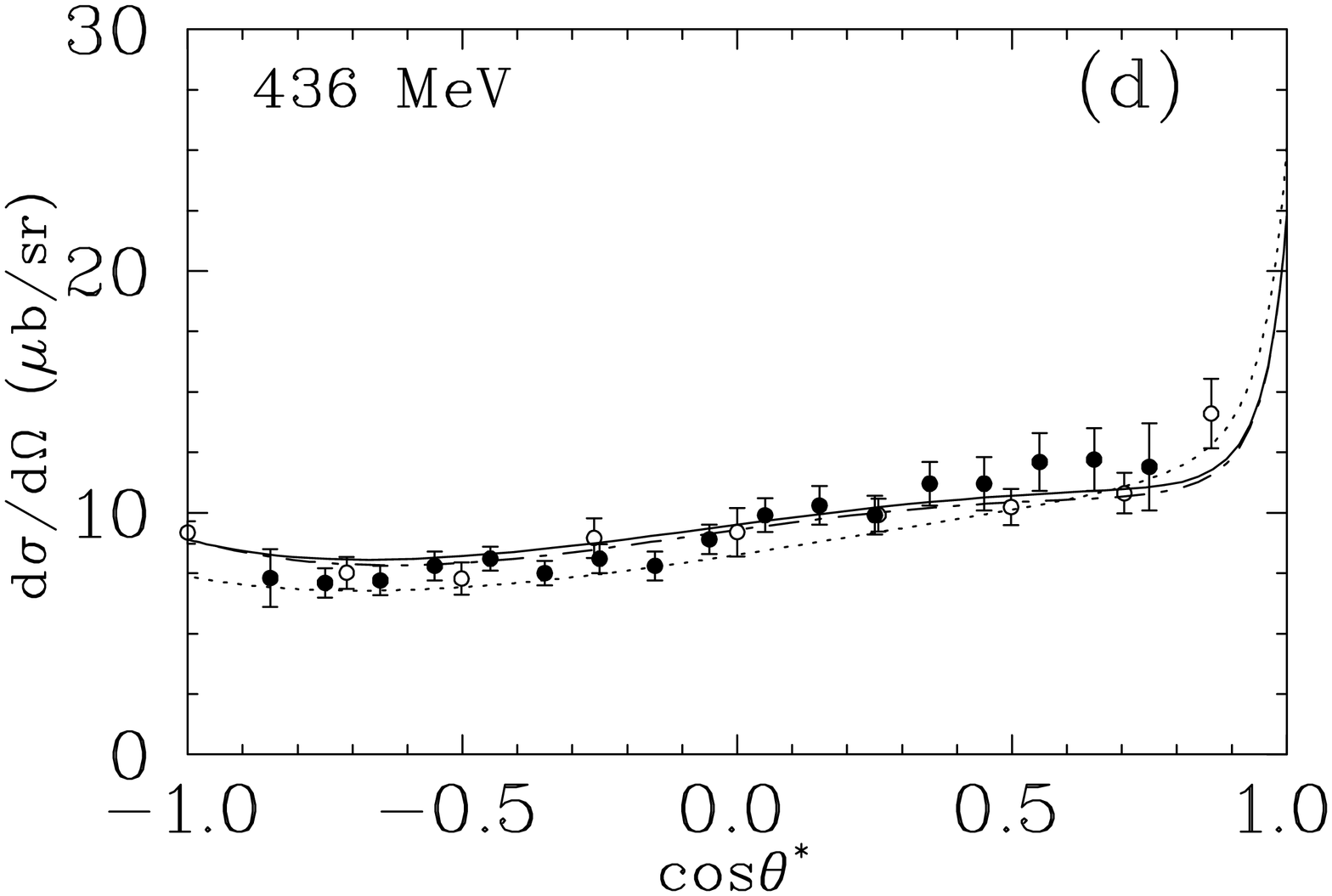}
\includegraphics[height=0.45\textwidth, angle=90]{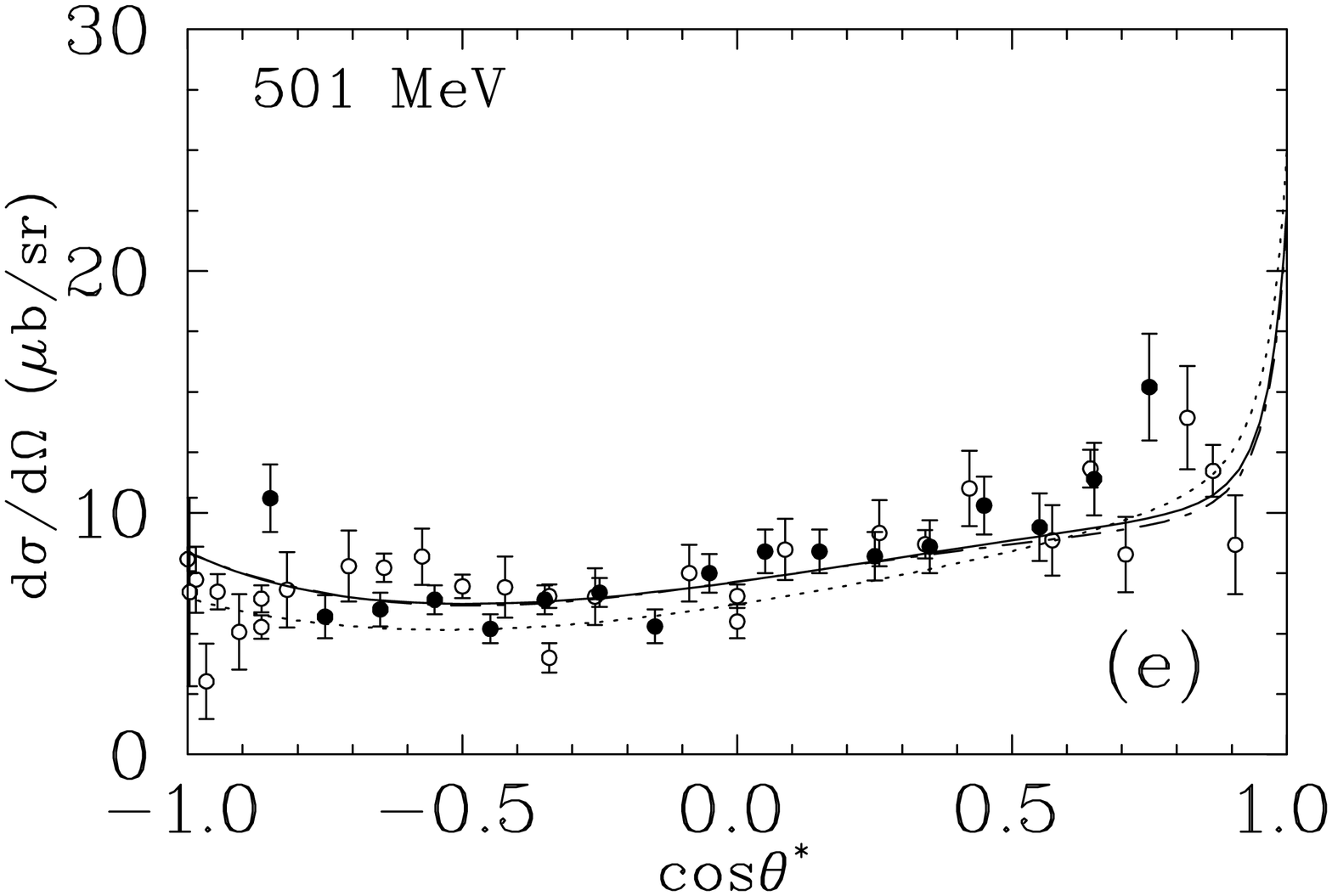}\hfill
\includegraphics[height=0.45\textwidth, angle=90]{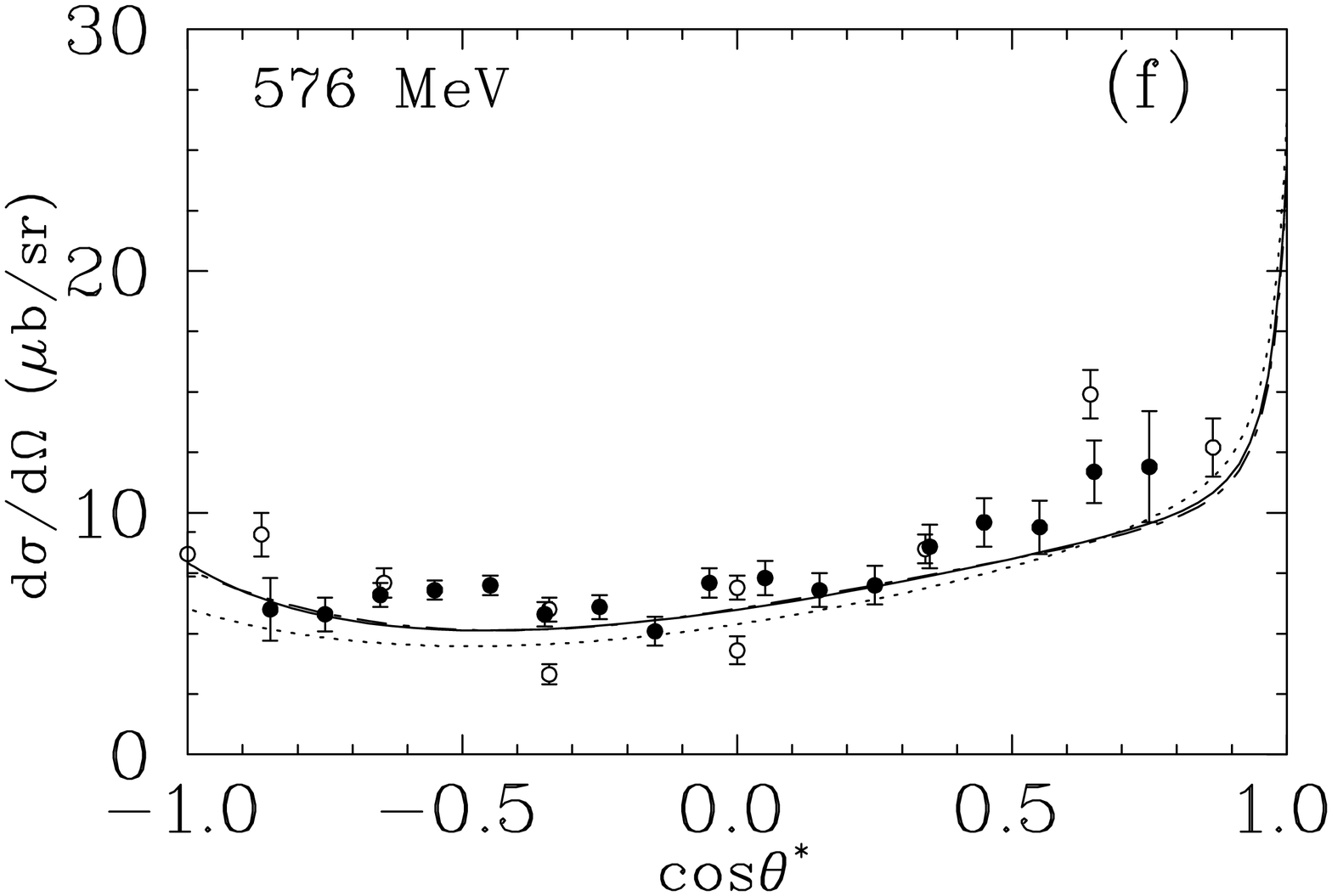}
\includegraphics[height=0.45\textwidth, angle=90]{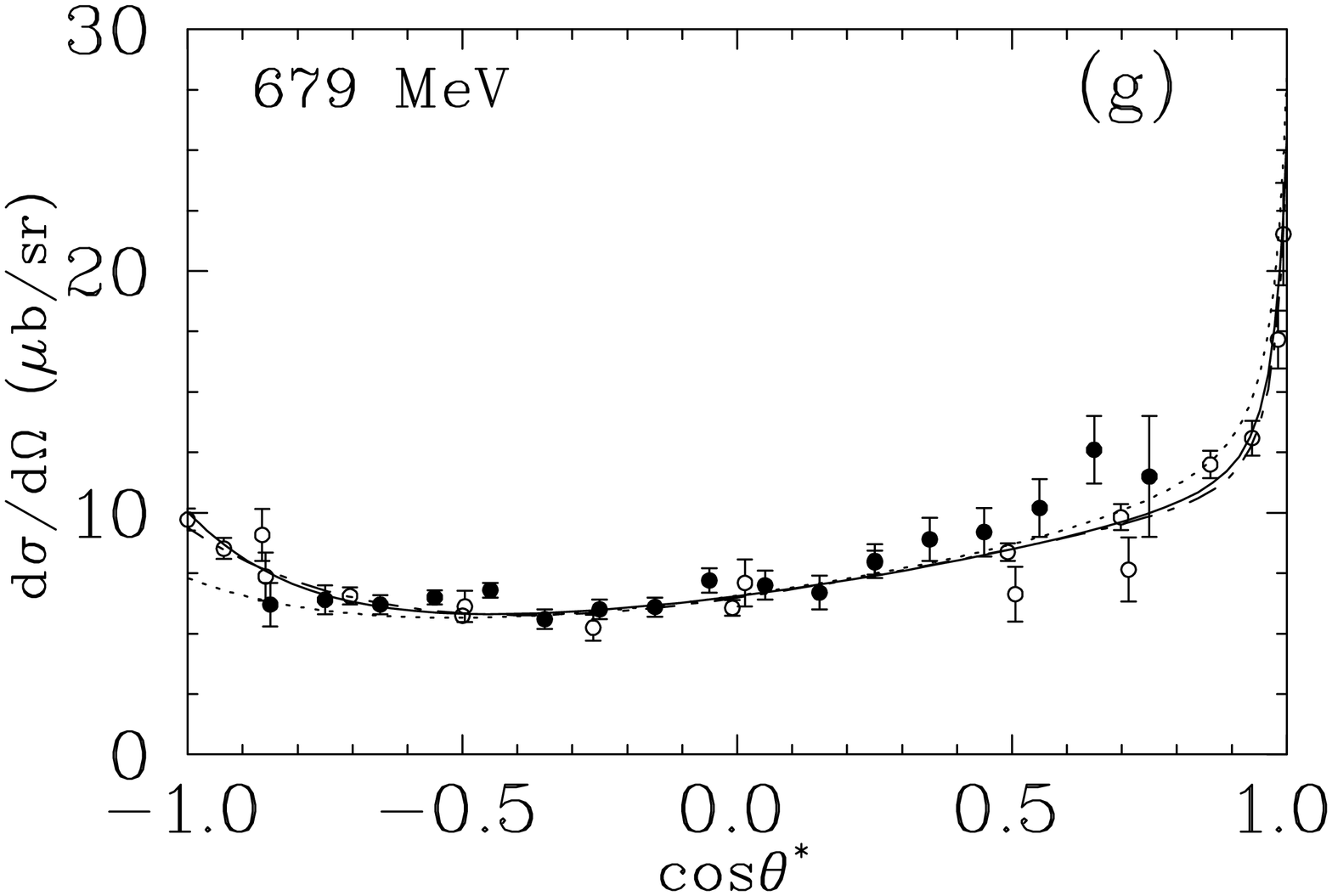}\hfill
\includegraphics[height=0.45\textwidth, angle=90]{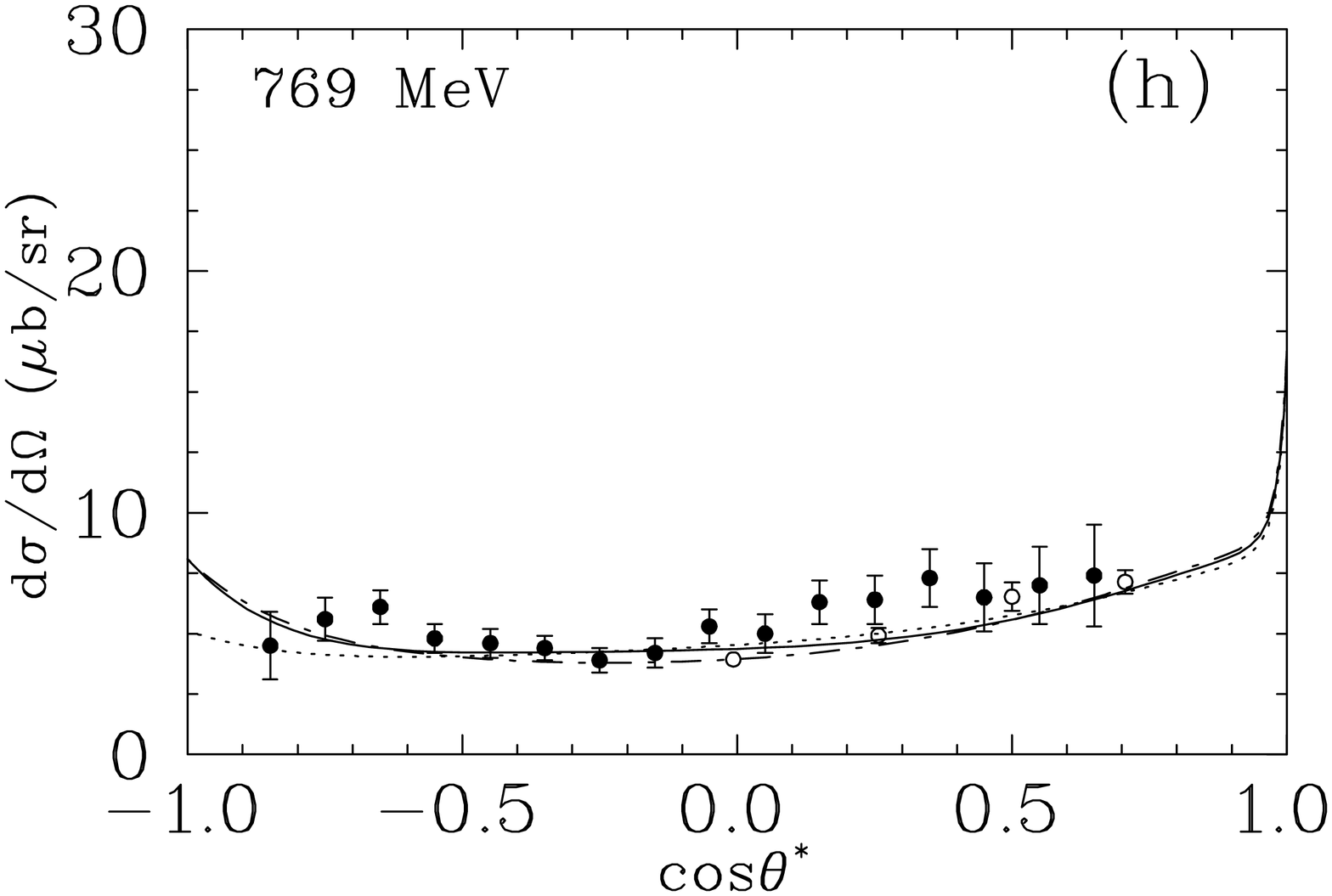}
}\caption{Differential cross sections for $\gamma
          n\to\pi^-p$ at (a) 285~MeV, (b) 338~MeV,
          (c) 407~MeV, (d) 436~MeV, (e) 501~MeV,
          (f) 576~MeV, (g) 679~MeV, and (h) 769~MeV.
          The uncertainties are statistical
          only.  Dash-dotted (solid) curves
          correspond to the GW SAID SM02 (SH04)
          solution~\protect\cite{GWpr}.  The MAID
          solutions~\protect\cite{maid} are plotted with dashed lines.  Previous
          measurements~\protect\cite{bq73,ro73,REXMES} are shown as
          open circles. \label{fig:fi1}}
\end{figure*}

\begin{figure*}
\centering{
\includegraphics[height=0.45\textwidth, angle=90]{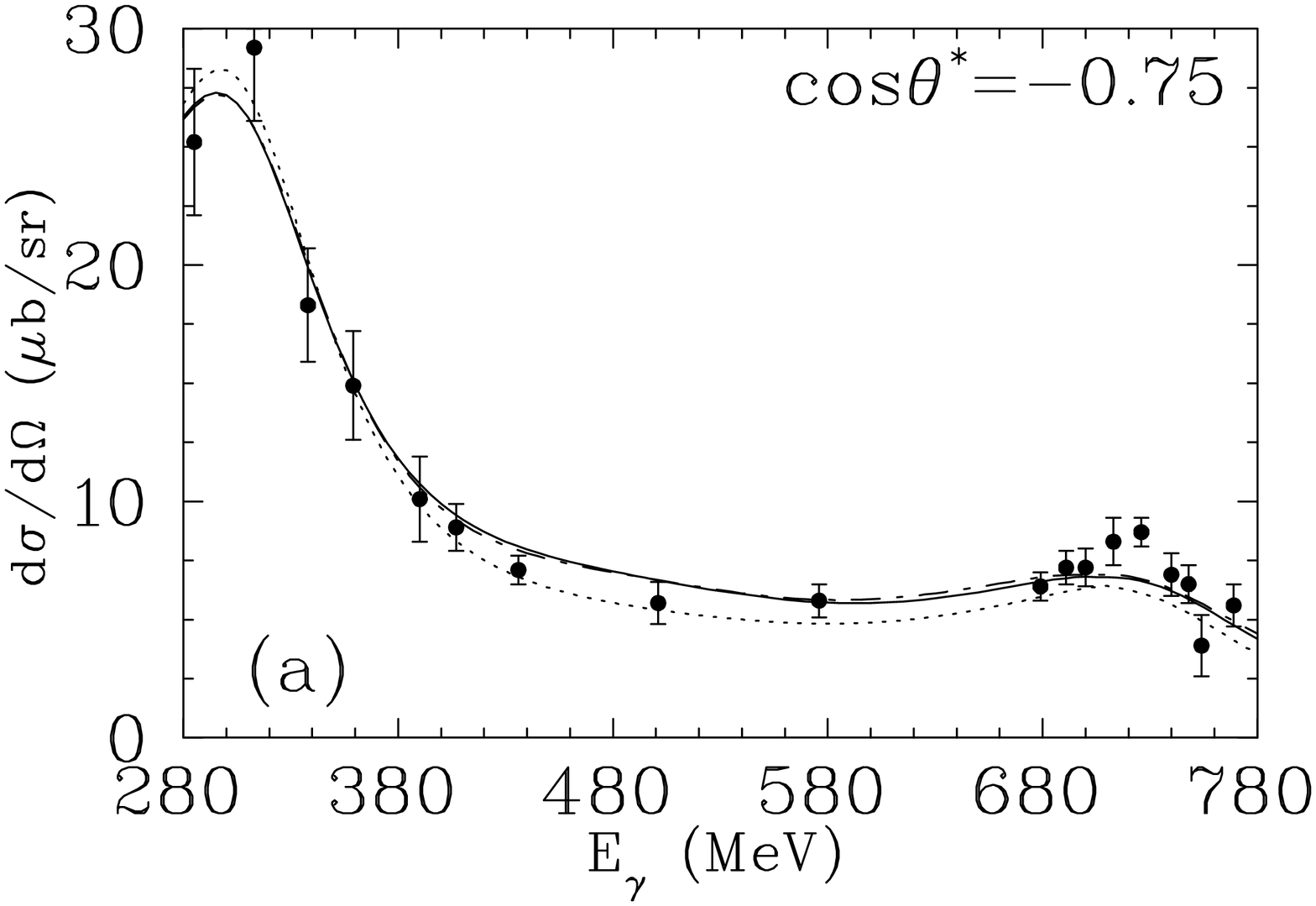}\hfill
\includegraphics[height=0.45\textwidth, angle=90]{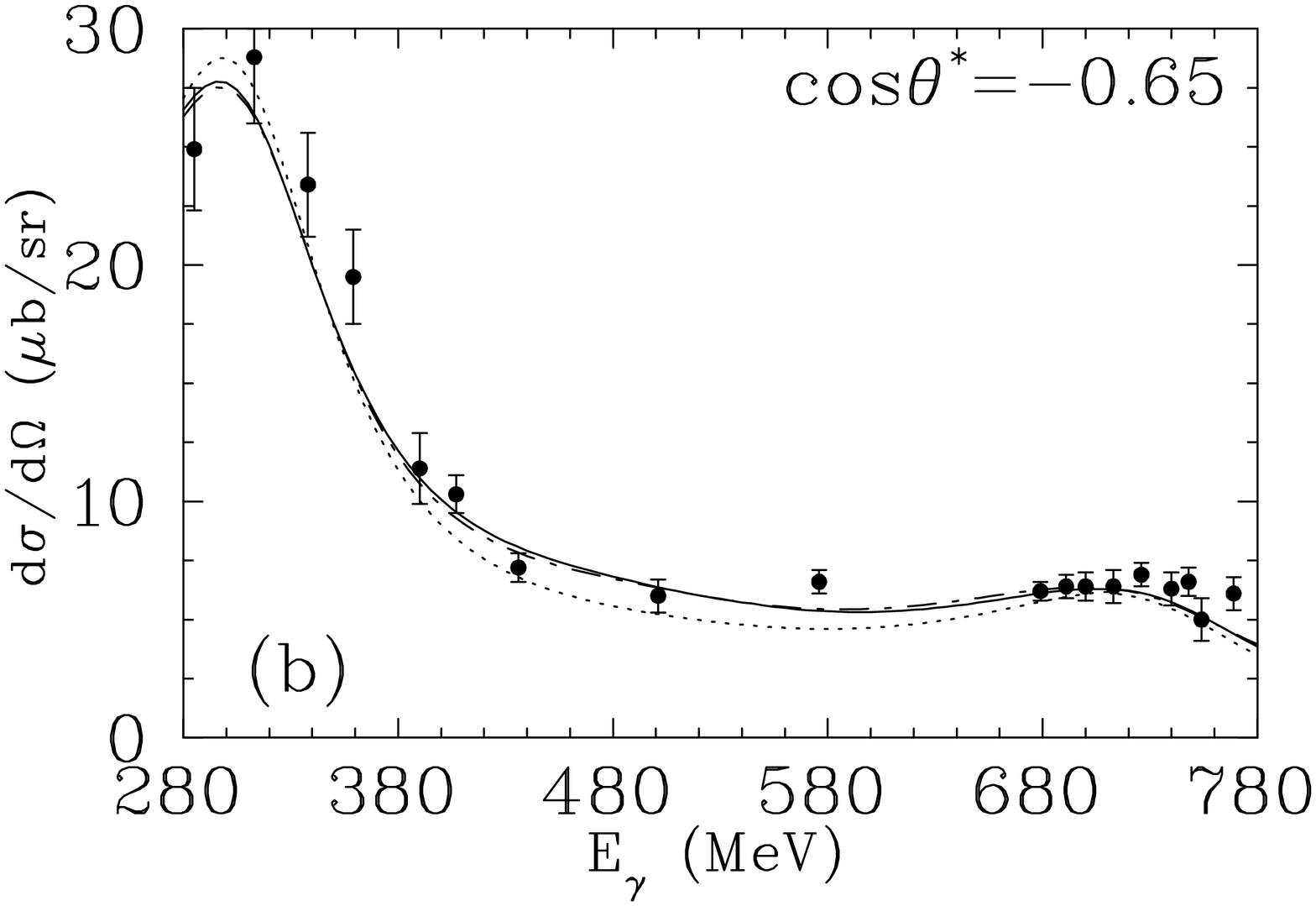}
\includegraphics[height=0.45\textwidth, angle=90]{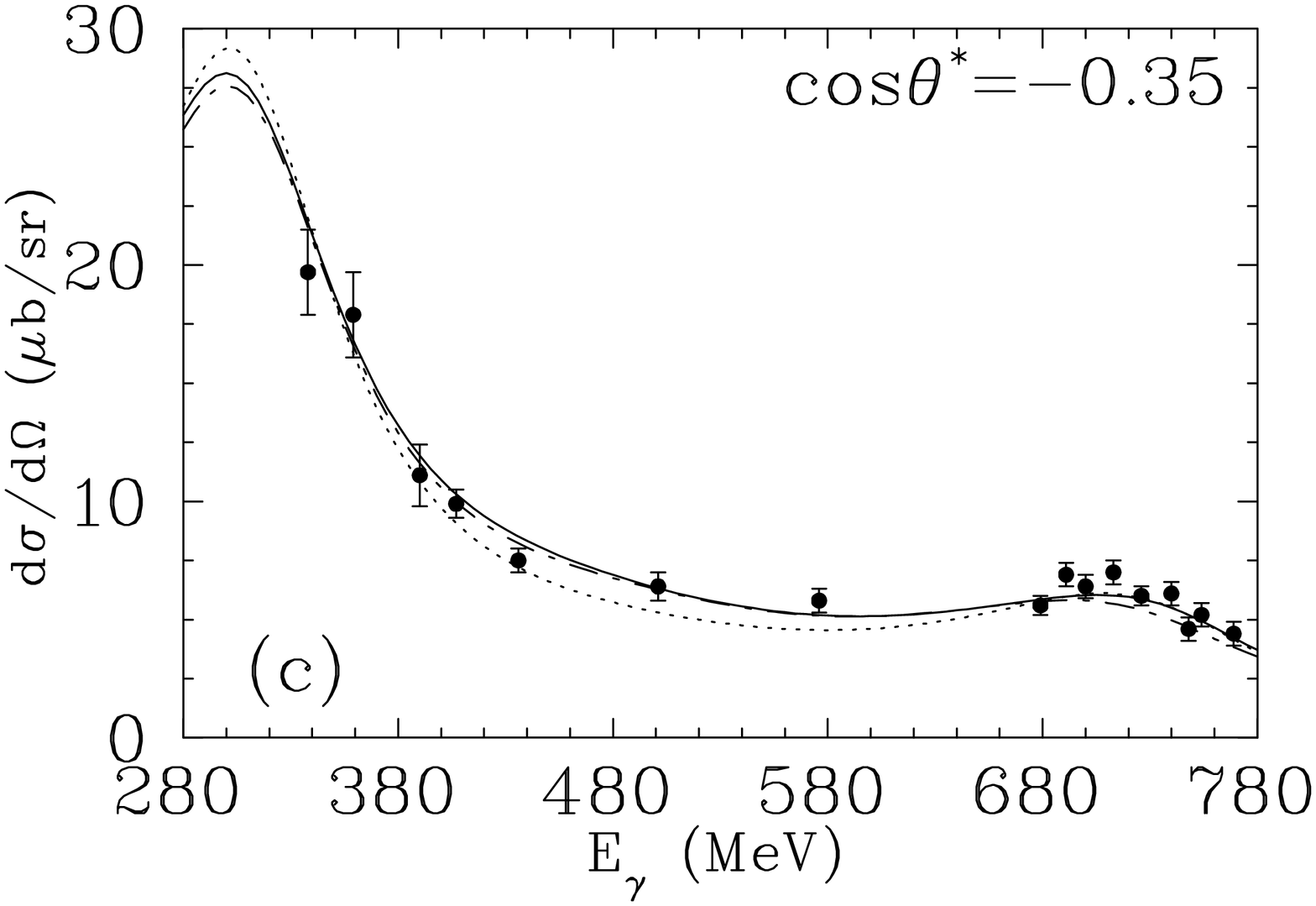}\hfill
\includegraphics[height=0.45\textwidth, angle=90]{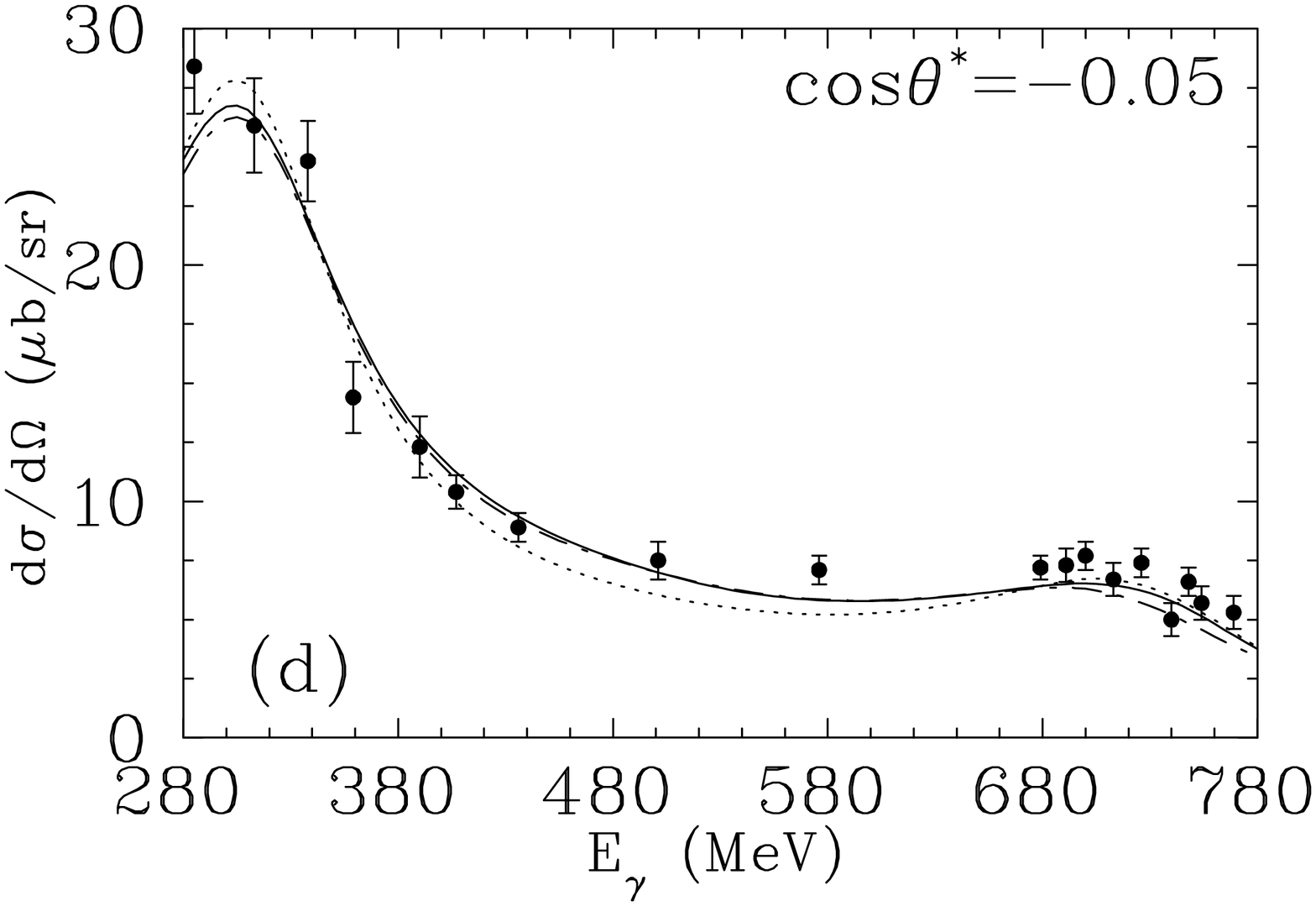}
\includegraphics[height=0.45\textwidth, angle=90]{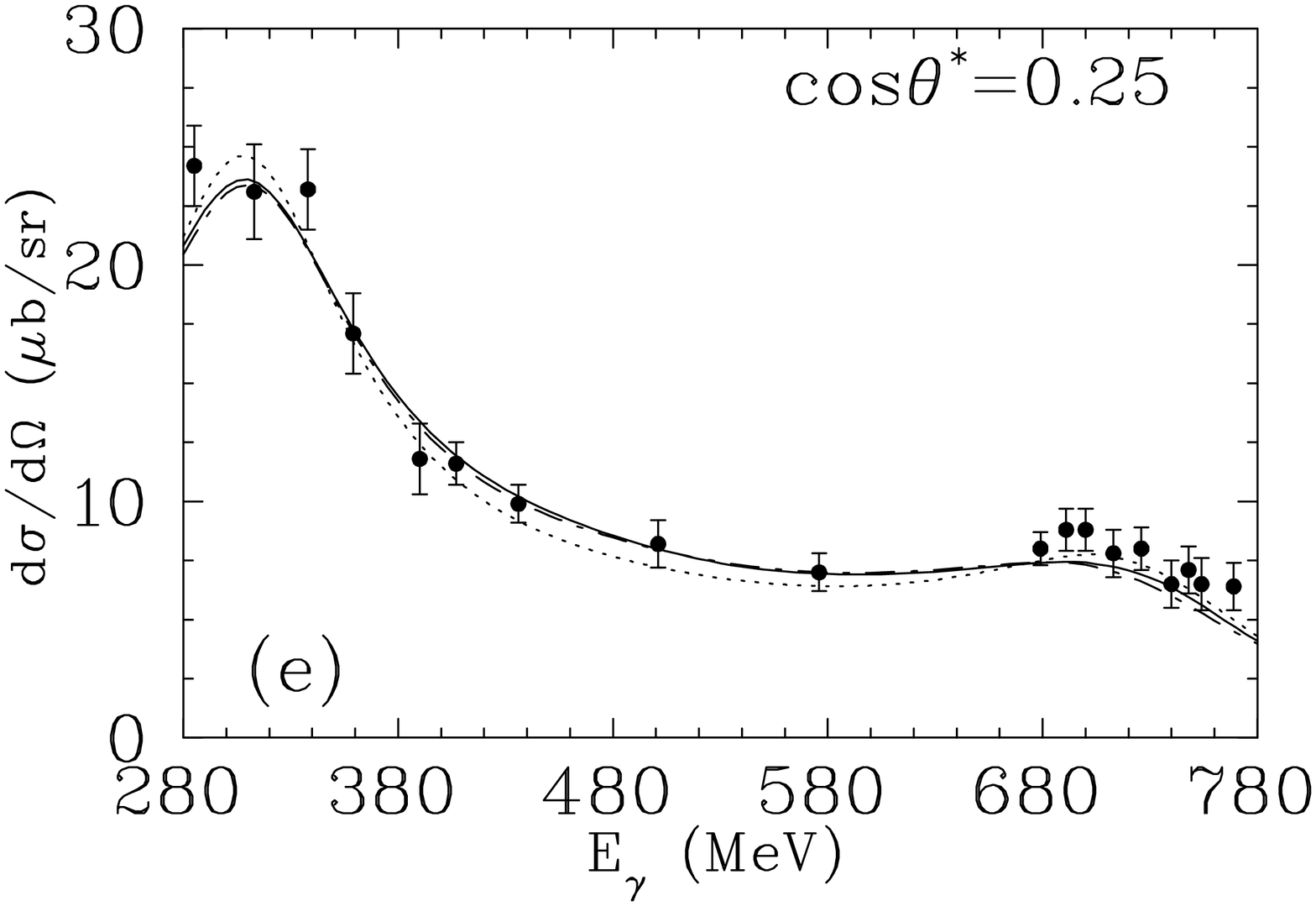}\hfill
\includegraphics[height=0.45\textwidth, angle=90]{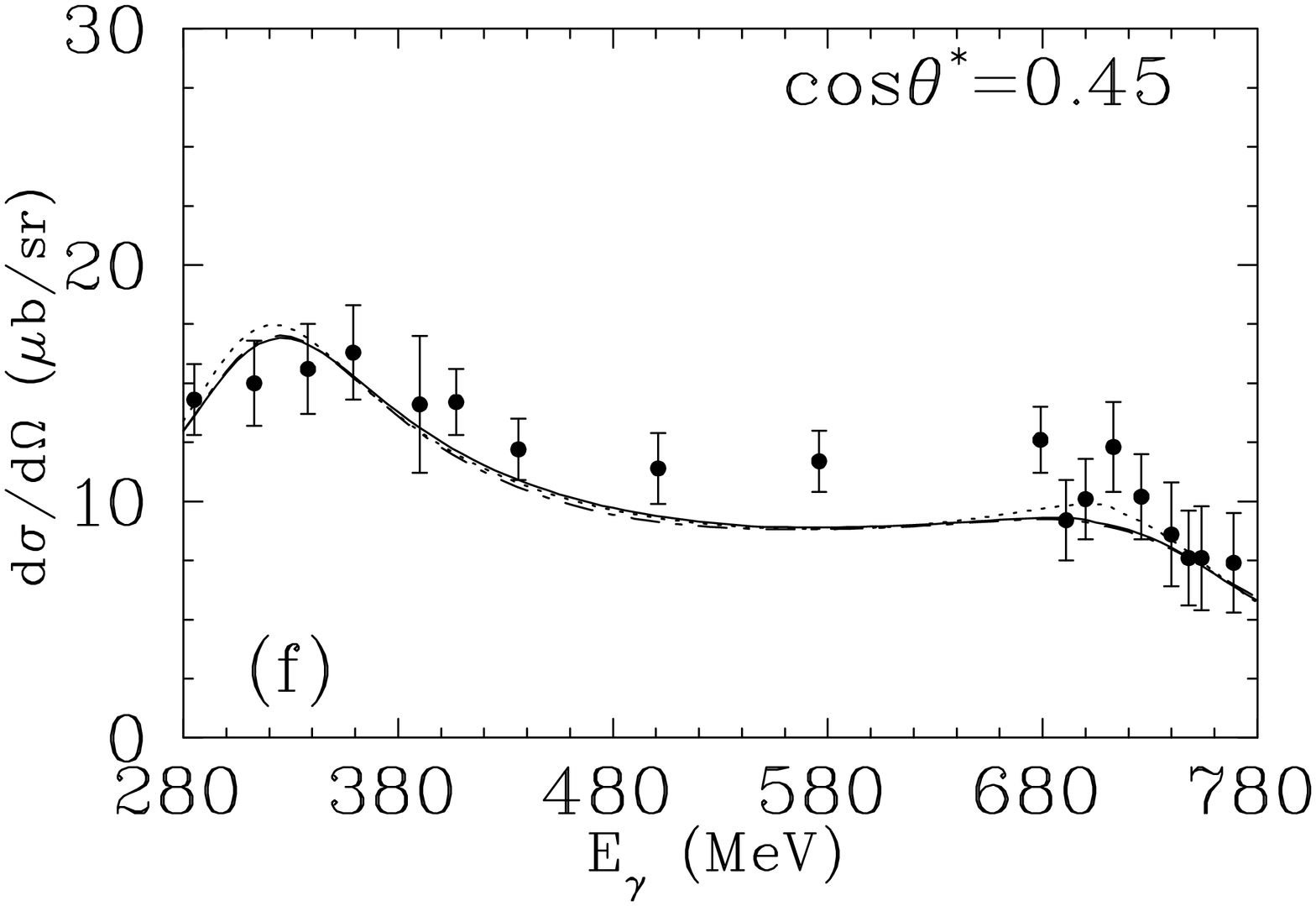}
}\caption{Differential cross sections for $\gamma
          n\to\pi^-p$ at (a) $\cos\theta^\ast = -0.75$,
          (b) $\cos\theta^\ast = -0.65$, (c) $\cos\theta^\ast =
          -0.35$, (d) $\cos\theta^\ast = -0.05$, (e)
          $\cos\theta^\ast = +0.25$, and (f) $\cos\theta^\ast =
          +0.45$.  Dash-dotted (solid) curve
          corresponds to the GW SM02 (SH04)
          solution~\protect\cite{GWpr}.  The MAID
          solution~\protect\cite{maid} predictions are
          plotted with dashed lines.  The quoted
          uncertainties are statistical only.
          \label{fig:fi2}}
\end{figure*}

Our results in the form $d\sigma/d\Omega(\gamma
n\rightarrow\pi^-p)$ are presented in Tables~\ref{tbl2},
\ref{tbl3}, and \ref{tbl4}.  A representative selection of angular
distributions is shown in Fig.~\ref{fig:fi1}.  Six examples of the
excitation function at cos($\theta^{\ast}$) = -0.75, -0.65, -0.35,
-0.05, 0.25, and 0.45 are shown in Fig.~\ref{fig:fi2}. The
enhancement at low $E_\gamma$ is due to the high-energy tail of
the $\Delta(1232)$, and the small bump at large $E_\gamma$
reflects the production of the $N(1520)$ and $N(1535)$.  The
excitation functions at all angles reveal no bump or shoulder that
could be indicative of the excitation of the Roper resonance. To
extract electromagnetic quantities for the Roper resonance, one
should use a multipole analysis of the available pion
photoproduction data. This is discussed in the next section.

The existing data come in two types: REX, as measured in our
experiment, and $\pi^-$-photoproduction on the deuteron. Our data
are more numerous (300 points) than any of the existing REX data
sets, generally agreeing with the photoproduction results that use
the so called ${\pi^-}/{\pi^+}$ technique for extracting the
$\pi^-$-photoproduction data from a deuterium target\cite{Aziz}.
Statistical uncertainties, which include angle-dependent
uncertainties due to background subtraction and acceptance
corrections, generally vary from 5\% to 15\%, except for the most
forward and backward scattering angles at low momenta where
statistical uncertainties are as large as 30\% for the
measurements reported in this paper. The data with larger
uncertainties at extreme scattering angles at each beam momentum
were eliminated if either the background subtraction was very
large or the acceptance was varying rapidly and the resulting
angle-dependent uncertainty is greater than 30\%. An overall
systematic uncertainty for all energy sets of about 5\% is
obtained, from the sum in quadrature of all other known factors.
For details, see Section~\ref{sec:distrib}.

%%%%%%%%%%%%%%%%%%%%%%%%%%%%%%%%%%%%%%%%%%
%%%%%%%%%%%%%%%%%%%%%%%%%%%%%%%%%%%%%%%%%%%

Figure~\ref{fig:fi1} also displays our comparison of the
predictions from the SAID PWA~\cite{GWpr} and the MAID~\cite{maid}
analyses of existing data. In both analyses, large disagreements
with some older bremsstrahlung measurements~\cite{bq73,ro73} are
obvious, particularly at lower energies.  Also plotted is the
best-fit (SH04), based on a SAID analysis incorporating the
current set of measurements.  Both the SAID and MAID predictions
give a good qualitative representation of the data. The CB data
and curves are presented without any renormalization. Upon
inclusion of the CB cross sections in the SH04 fit, the overall
$\chi^2$ dropped, relative to the SAID prediction, by only 25 (out
of 327).  We note that the structure near 700~MeV in the
excitation cross sections of Fig.~\ref{fig:fi2} appears sharper
than predicted by SAID and MAID. The reproduction of this feature
was not significantly improved in the SH04 fit.

%%%%%%%%%%%%%%%%%%%%%%%%%%%%%%%%%%%%%%%%%%%
%%%%%%%%%%%%%%%%%%%%%%%%%%%%%%%%%%%%%%%%%%%

\begin{figure*}
\centering{
\includegraphics[height=0.5\textwidth, angle=90]{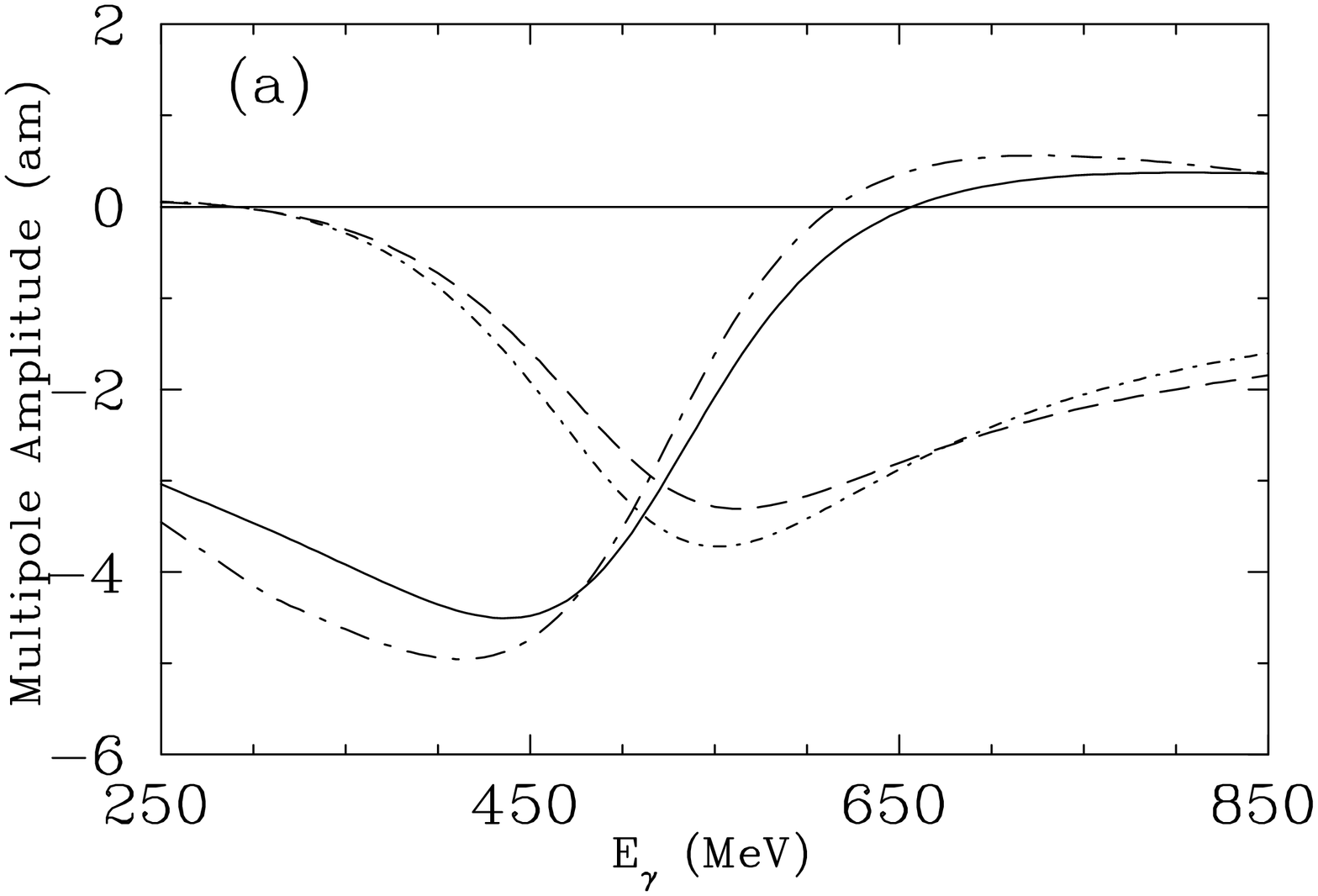}\hfill
\includegraphics[height=0.5\textwidth, angle=90]{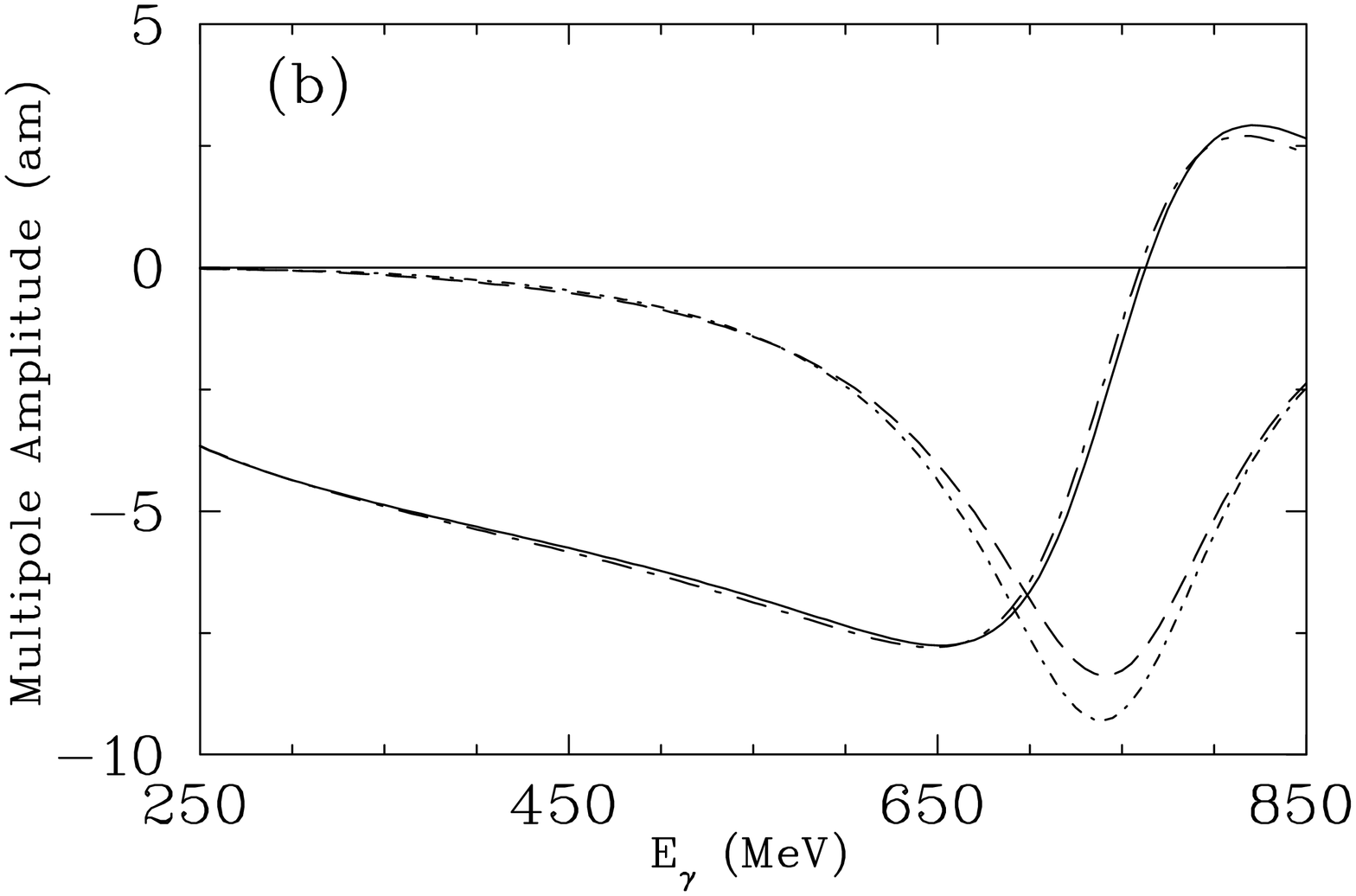}
}\caption{Major multipoles affected by the CB data:
         (a) $\rm _nM_{1-}^{1/2}$ and (b) $\rm
          _nE_{2-}^{1/2}$.  Solid (dashed) curves
         give the real (imaginary) parts of
         amplitudes corresponding to the GW SM02
         solution~\protect\cite{GWpr}.  The real
         (imaginary) parts of GW SH04 solution,
         are plotted as dash-dotted (short
         dash-dotted) curves.  The subscript n
         denotes a neutron target.
         \label{fig:new}}
\end{figure*}

The new CB cross sections did not result in large changes to the
multipole amplitudes. Examples of multipoles on a neutron target
showing typical differences are plotted in Fig.~\ref{fig:new}.
However, when we include the resulting multipoles in a new
three-parameter fit to extract the $A^n_{\frac{1}{2}}$ amplitude
for the $P_{11}$, a reduction from $47\pm5$ to $36\pm7$
(GeV)$^{-1/2}\times 10^{-3}$ is seen in the $A^n_{\frac{1}{2}}$
amplitude with the uncertainty in the amplitude being mainly due
to the non-resonant background. This value for $A^n_{\frac{1}{2}}$
agrees very well with the value currently quoted by the PDG. While
these analyses were performed including all published data, we did
experiment with the fits by removing older data that were more
than 5 standard deviations from the new fit. While this process
decreased the reduced $\chi^2$ to $\sim1.0$, all other results
remained the same.

%%%%%%%%%%%%%%%%

%%%%%%%%%%%%%%%%%%%%%%%%%%%%%%%%%%%%%%%%%%%%%%%%%%%%%
%%%  IV. Conclusion and discussion
%%%%%%%%%%%%%%%%%%%%%%%%%%%%%%%%%%%%%%%%%%%%%%%%%%%%%
\section{Conclusion and discussion}
\label{sec:summ}

We have measured a comprehensive set of differential cross
sections at 18 energies for the inverse pion photoproduction
reaction, $\pi^-p\to n\gamma$, over an energy range most sensitive
to contributions from the Roper resonance. As is evident in
Fig.~\ref{fig:fi1}, the existing database for this reaction was
populated by a number of measurements inconsistent with the
extensive SAID and MAID fits to both neutron and proton target
data. Our measurements have verified the features of both the SAID
and MAID analyses, however they disagree with some of the older
data sets. A major accomplishment of this experiment is a
substantial improvement in the $\pi^-$-photoproduction data base,
adding 300 new differential cross sections. Inclusion of these new
data has resulted in only small changes to the SAID multipole
amplitudes, but a drop in the value of $A^n_{\frac{1}{2}}$ was
seen.

The lowest-energy differential cross section angular distribution
in Fig.~\ref{fig:fi1} shows that the SAID and MAID fits are in
disagreement with much of the older data, but are in satisfactory
agreement with the current measurement. This observation reflects
the fact that neither fit is entirely model-independent and
individual data sets do not determine fits. Both fits use similar
prescriptions to unitarize the Born-term background. In the MAID
approach, resonances are added explicitly; in SAID, resonance
contributions are added implicitly through a parameterization in
terms of the $\pi N$ T-matrix. This limitation on the form does
not allow a fit to arbitrary angular variation, especially at
lower energies, such as the backward dip suggested by the older
data at 285~MeV as is shown very clearly in Fig.~\ref{fig:fi1}a:
the SAID and MAID solutions "predicted" our cross section values
at 285 MeV despite the low values of the older data.

The largest differences between SAID and MAID are visible at
forward and backward angles in Fig.~\ref{fig:fi1}, at the highest
energies. Disagreements between the analyses and the new CB data
are also enhanced in these regions, as we have shown in
Fig.~\ref{fig:fi2}. It should be emphasized that the differences
in the multipoles are not major and indicate that the data base is
approaching an accurate representation of the REX interaction over
the energy range that covers the Roper resonance, and that one can
certainly rely on the SAID and MAID representations at the 10\%
level. We have obtained a reduction in the $A^n_{\frac{1}{2}}$
amplitude for the $P_{11}$ (from $47\pm5$ to $36\pm7$
(GeV)$^{-1/2}\times 10^{-3}$) when we include our new multipoles
in a three parameter fit. This value agrees with the value quoted
in Table I for the PDG. The largest uncertainty in the amplitude
extraction is due the method used to handle the non-resonant
background in this fit. The remaining differences in the neutron
and proton couplings for the Roper resonance in Table I are
speculated to be due to the different extraction methods used by
each author. These differences must be settled by further
theoretical work.

On the experimental side, further improvements in the PWAs await
more data in the region at and above the $N(1535)$ where the
number of measurements for this reaction is small. Of particular
importance in all energy regions is the need for data obtained
involving polarized photons and polarized targets. Due to the
closing of hadron facilities, new $\pi^-p \rightarrow \gamma n $
experiments are not in the planning and only $\gamma n \rightarrow
\pi^- p  $ measurements are possible at electron facilities using
deuterium or helium targets. Our agreement with the existing
$\pi^-$ photoproduction reaction measurements lead us to believe
that the photoproduction measurements are reliable despite the
necessity of using a deuterium target. Plans are now in place to
use the Crystal Ball and the polarized photon beam and polarized
$^1$H, $^2$H, and $^3$He targets at MAMI C to collect
photoproduction data off the neutron in the next several years.

%%%%%%%%%%%%%%%%%%%%%%%%%%%%%%%%%%%%%%%%%%%%%%%%%%%%%%%%%
%%%   Acknowledgments
%%%%%%%%%%%%%%%%%%%%%%%%%%%%%%%%%%%%%%%%%%%%%%%%%%%%%%%%%
\acknowledgments We thank A.~E.~Kudryavtsev and A.~Donnachie for
useful discussions on $\gamma n$ scattering.  The authors
acknowledge the support of the US National Science Foundation, the
US Department of Energy, NSERC of Canada, the Russian Ministry of
Science and Technology, the Russian Foundation of Basic Research,
the Russian State Scientific-Technical Program: Fundamental
Nuclear Physics, the Croatian Ministry of Science, and the George
Washington University Research Enhancement Fund. We thank SLAC for
the loan of the Crystal Ball. The assistance of the staff of BNL
and AGS is much appreciated.

%%%%%%%%%%%%%%%%%%%%%%%%%%%%%%%%%%%%%%%%%%%%%%%%%%%%%%%%%
%%%    V. References
%%%%%%%%%%%%%%%%%%%%%%%%%%%%%%%%%%%%%%%%%%%%%%%%%%%%%%%%%

\end{document}